\newcommand{\be}{\begin{equation}}
\newcommand{\bea}{\begin{eqnarray}}
\newcommand{\ee}{\end{equation}}
\newcommand{\eea}{\end{eqnarray}}

\newcommand{\qa}{\alpha}

\newcommand{\qd}{\delta}

\newcommand{\qy}{\theta}

\newcommand{\qr}{\rho}
\newcommand{\qs}{\sigma}

\newcommand{\qf}{\varphi}
\newcommand{\qF}{\Phi}

\newcommand{\qj}{\psi}
\newcommand{\qJ}{\Psi}

\newcommand{\tr}{{\rm tr}\,}

\newcommand{\dagg}{^{\dag}}

\newcommand{\fr}[2]{{\textstyle \frac{#1}{#2}}}

\newcommand{\bits}{ \{0,1\} }

\newcommand{\cD}{{\mathcal D}}

\newcommand{\cH}{{\mathcal H}}

\newcommand{\cM}{{\mathcal M}}

\newcommand{\cX}{{\mathcal X}}

\newcommand{\pr}{{\rm Pr}}

\newcommand{\isdef}{\stackrel{\rm def}{=}}

\newcommand{\ket}[1]{| #1 \rangle}
\newcommand{\bra}[1]{\langle #1 |}
\newcommand{\inprod}[2]{\langle #1 | #2 \rangle}

\newtheorem{theorem}{Theorem}
\newtheorem{lemma}{Lemma}

\documentclass[]{article}

\usepackage{tikzit}
\usepackage{amsmath}
\usepackage{amsfonts}
\usepackage{mathtools}
\usepackage{amssymb}
\usepackage{gensymb}
\usepackage{hyperref}
\usepackage{xcolor}
\usepackage[margin=1.25in]{geometry}
\usepackage{blochsphere}
\newcommand{\equaltext}[1]{\ensuremath{\stackrel{\text{#1}}{=}}}

\usepackage{tikz}
\usetikzlibrary{backgrounds}
\usetikzlibrary{arrows}
\usetikzlibrary{shapes,shapes.geometric,shapes.misc}

\tikzstyle{every picture}=[baseline=-0.25em,scale=0.5]

\pgfkeys{/tikz/tikzit fill/.initial=0}
\pgfkeys{/tikz/tikzit draw/.initial=0}
\pgfkeys{/tikz/tikzit shape/.initial=0}
\pgfkeys{/tikz/tikzit category/.initial=0}

\input{computation.tikzdefs}

\tikzstyle{medium box}=[fill=white, draw=black, shape=rectangle, minimum width=1cm, minimum height=0.6cm]
\tikzstyle{MEDIUM BOX}=[fill=white, draw=black, shape=rectangle, minimum width=1cm, minimum height=0.6cm, line width=1]
\tikzstyle{asym box}=[shape=NWbox, minimum width=0.2cm, minimum height=0.6cm, tikzit shape=rectangle, tikzit fill=red, draw=black, fill=white, font={\footnotesize}, xscale=-1]
\tikzstyle{ASYM BOX}=[shape=NWbox, minimum width=0.2cm, minimum height=0.6cm, tikzit shape=rectangle, tikzit fill=red, draw=black, fill=white, font={\footnotesize}, xscale=-1, line width=1]
\tikzstyle{asym box conjugate}=[shape=NWbox, minimum width=0.2cm, minimum height=0.6cm, tikzit shape=rectangle, tikzit fill=red, draw=black, fill=white, font={\footnotesize}]
\tikzstyle{asym box transpose}=[shape=SWbox, minimum width=0.2cm, minimum height=0.6cm, tikzit shape=rectangle, tikzit fill=red, draw=black, fill=white, font={\footnotesize}]
\tikzstyle{asym box adjoint}  =[shape=SEbox, minimum width=0.2cm, minimum height=0.6cm, tikzit shape=rectangle, tikzit fill=red, draw=black, fill=white, font={\footnotesize}]
\tikzstyle{asym mid}=[shape=Wpara, minimum width=0.2cm, tikzit shape=rectangle, tikzit fill={rgb,255: red,128; green,0; blue,128}, fill=white, draw=black, minimum height=0.6cm, font={\footnotesize}]
\tikzstyle{StateLeft}=[shape=SEflag, minimum width=0.6cm, tikzit shape=rectangle, tikzit fill=blue, fill=white, draw=black, minimum height=0.6cm, font={\footnotesize}, rotate=180]
\tikzstyle{EffectRight}=[fill=white, draw=black, shape=SEflag, rotate=0, tikzit shape=rectangle, tikzit fill={rgb,255: red,255; green,191; blue,191}, minimum height=0.6cm, minimum width=0.6cm]
\tikzstyle{EFFECTRIGHT}=[fill=white, draw=black, shape=SEflag, rotate=0, tikzit shape=rectangle, tikzit fill={rgb,255: red,255; green,191; blue,191}, minimum height=0.6cm, line width=1, minimum width=0.6cm]
\tikzstyle{StateRight}    =[shape=SEflag, tikzit shape=rectangle, tikzit fill={rgb,255: red,0; green,128; blue,128}, fill=white, draw=black, font={\footnotesize}, rotate=270, minimum width=0.6cm, minimum height=0.6cm]
\tikzstyle{STATERIGHT}    =[shape=SEflag, tikzit shape=rectangle, tikzit fill={rgb,255: red,0; green,128; blue,128}, fill=white, draw=black, font={\footnotesize}, rotate=270, minimum width=0.6cm, line width=1, minimum height=0.6cm]
\tikzstyle{WideStateRight}=[shape=SEflag, tikzit shape=rectangle, tikzit fill={rgb,255: red,0; green,128; blue,128}, fill=white, draw=black, font={\footnotesize}, rotate=270, minimum width=0.6cm,  minimum height=1.2cm]
\tikzstyle{WIDESTATERIGHT}=[shape=SEflag, tikzit shape=rectangle, tikzit fill={rgb,255: red,0; green,128; blue,128}, fill=white, draw=black, font={\footnotesize}, rotate=270, minimum width=0.6cm, line width=1, minimum height=1.2cm]
\tikzstyle{WIDEEFFECTRIGHT}=[shape=SEflag, tikzit shape=rectangle, tikzit fill={rgb,255: red,0; green,128; blue,128}, fill=white, draw=black, font={\footnotesize}, rotate=90, minimum width=0.6cm, line width=1, minimum height=1.2cm, yscale=-1]
\tikzstyle{EffectMid}=[shape=SEflag, minimum width=0.6cm, tikzit shape=rectangle, tikzit fill=white, fill=white, draw=black, minimum height=0.6cm, font={\footnotesize}, rotate=225, left=-0.1285cm, label={[shift={(-0.45,0.5)}]${}$}, yshift=-0.1285cm, tikzit category=mid]
\tikzstyle{EFFECTMID}=[shape=SEflag, minimum width=0.6cm, tikzit shape=rectangle, tikzit fill=white, fill=white, draw=black, minimum height=0.6cm, line width=1, font={\footnotesize}, rotate=225, left=-0.1285cm, label={[shift={(-0.45,0.5)}]${}$}, yshift=-0.1285cm, tikzit category=mid]
\tikzstyle{EffectLeft}=[fill=white, draw=black, shape=SEflag, rotate=90, tikzit shape=rectangle, tikzit fill=yellow, minimum height=0.6cm, minimum width=0.6cm]
\tikzstyle{WhiteSpider}=[fill=white, draw=black, shape=circle, minimum size=0.2cm, inner sep=0 cm]
\tikzstyle{GraySpider}=[fill={rgb,255: red,191; green,191; blue,191}, draw=black, shape=circle, minimum size=0.2cm, inner sep=0 cm]
\tikzstyle{WHITESPIDER}=[fill=white, draw=black, shape=circle, line width=1, minimum size=0.2cm, inner sep=0 cm]
\tikzstyle{GRAYSPIDER}=[fill={rgb,255: red,191; green,191; blue,191}, draw=black, shape=circle, minimum size=0.2cm, inner sep=0 cm, line width=1]
\tikzstyle{StateMidGray}=[shape=SEflag, minimum width=0.6cm, tikzit shape=rectangle, tikzit fill={rgb,255: red,191; green,191; blue,191}, fill={rgb,255: red,191; green,191; blue,191}, draw=black, minimum height=0.6cm, font={\footnotesize}, rotate=45, left=-0.1285cm, label={[shift={(0.45,-0.50)}]${}$}, yshift=-0.1285cm, tikzit category=mid]
\tikzstyle{STATEMIDGRAY}=[shape=SEflag, minimum width=0.6cm, tikzit shape=rectangle, tikzit fill={rgb,255: red,191; green,191; blue,191}, fill={rgb,255: red,191; green,191; blue,191}, draw=black, minimum height=0.6cm, font={\footnotesize}, rotate=45, left=-0.1285cm, label={[shift={(0.45,-0.50)}]${}$}, yshift=-0.1285cm, tikzit category=mid, line width=1]
\tikzstyle{EffectMidGray}=[shape=SEflag, minimum width=0.6cm, tikzit shape=rectangle, tikzit fill={rgb,255: red,191; green,191; blue,191}, fill={rgb,255: red,191; green,191; blue,191}, draw=black, minimum height=0.6cm, font={\footnotesize}, rotate=225, left=-0.1285cm, label={[shift={(0.45,-0.50)}]${}$}, yshift=-0.1285cm, tikzit category=mid]
\tikzstyle{StateMid}=[shape=SEflag, minimum width=0.6cm, tikzit shape=rectangle, tikzit fill=white, fill=white, draw=black, minimum height=0.6cm, font={\footnotesize}, rotate=45, left=-0.1285cm, label={[shift={(0.45,-0.50)}]${}$}, yshift=-0.1285cm, tikzit category=mid]
\tikzstyle{StateMidBrown}=[shape=SEflag, minimum width=0.6cm, tikzit shape=rectangle, tikzit fill={rgb,255: red,191; green,128; blue,64}, fill=brown, draw=black, minimum height=0.6cm, font={\footnotesize}, rotate=45, left=-0.1285cm, label={[shift={(0.45,-0.50)}]${}$}, yshift=-0.1285cm, tikzit category=mid]
\tikzstyle{StateMidGreen}=[shape=SEflag, minimum width=0.6cm, tikzit shape=rectangle, tikzit fill=green, fill=green, draw=black, minimum height=0.6cm, font={\footnotesize}, rotate=45, left=-0.1285cm, label={[shift={(0.45,-0.50)}]${}$}, yshift=-0.1285cm, tikzit category=mid]
\tikzstyle{StateMidPink}=[shape=SEflag, minimum width=0.6cm, tikzit shape=rectangle, tikzit fill=pink, fill=pink, draw=black, minimum height=0.6cm, font={\footnotesize}, rotate=45, left=-0.1285cm, label={[shift={(0.45,-0.50)}]${}$}, yshift=-0.1285cm, tikzit category=mid]
\tikzstyle{StateMidBlue}=[shape=SEflag, minimum width=0.6cm, tikzit shape=rectangle, tikzit fill=blue, fill={blue!80}, draw=black, minimum height=0.6cm, font={\footnotesize}, rotate=45, left=-0.1285cm, label={[shift={(0.45,-0.50)}]${}$}, yshift=-0.1285cm, tikzit category=mid]
\tikzstyle{STATEMIDBROWN}=[shape=SEflag, minimum width=0.6cm, tikzit shape=rectangle, tikzit fill={rgb,255: red,191; green,128; blue,64}, fill=brown, draw=black, minimum height=0.6cm, font={\footnotesize}, rotate=45, left=-0.1285cm, label={[shift={(0.45,-0.50)}]${}$}, yshift=-0.1285cm, tikzit category=mid, line width=1]
\tikzstyle{STATEMIDGREEN}=[shape=SEflag, minimum width=0.6cm, tikzit shape=rectangle, tikzit fill=green, fill=green, draw=black, minimum height=0.6cm, font={\footnotesize}, rotate=45, left=-0.1285cm, label={[shift={(0.45,-0.50)}]${}$}, yshift=-0.1285cm, tikzit category=mid, line width=1]
\tikzstyle{STATEMIDPINK}=[shape=SEflag, minimum width=0.6cm, tikzit shape=rectangle, tikzit fill=purp, fill=pink, draw=black, minimum height=0.6cm, font={\footnotesize}, rotate=45, left=-0.1285cm, label={[shift={(0.45,-0.50)}]${}$}, yshift=-0.1285cm, tikzit category=mid, line width=1]
\tikzstyle{STATEMIDBLUE}=[shape=SEflag, minimum width=0.6cm, tikzit shape=rectangle, tikzit fill=blue, fill={blue!80}, draw=black, minimum height=0.6cm, font={\footnotesize}, rotate=45, left=-0.1285cm, label={[shift={(0.45,-0.50)}]${}$}, yshift=-0.1285cm, tikzit category=mid, line width=1]
\tikzstyle{STATEMID}=[shape=SEflag, minimum width=0.6cm, tikzit shape=rectangle, tikzit fill=white, fill=white, draw=black, minimum height=0.6cm, font={\footnotesize}, rotate=45, left=-0.1285cm, label={[shift={(0.45,-0.45)}]${}$}, yshift=-0.1285cm, tikzit category=mid, line width=1]
\tikzstyle{RedSpider}=[fill={red!80}, draw=black, shape=circle, minimum size=0.2cm, inner sep=0 cm]
\tikzstyle{GreenSpider}=[fill=green, draw=black, shape=circle, minimum size=0.2cm, inner sep=0 cm]
\tikzstyle{yellowSpider}=[fill=yellow, draw=black, shape=circle, minimum size=0.2cm, inner sep=0 cm]
\tikzstyle{BlueSpider}=[fill={blue!80}, draw=black, shape=circle, minimum size=0.2cm, inner sep=0 cm]
\tikzstyle{STATEMIDRED}=[shape=SEflag, minimum width=0.6cm, tikzit shape=rectangle, tikzit fill=red, fill={red!80}, draw=black, minimum height=0.6cm, font={\footnotesize}, rotate=45, left=-0.1285cm, label={[shift={(0.45,-0.45)}]${}$}, yshift=-0.1285cm, tikzit category=mid, line width=1]
\tikzstyle{EFFECTMIDRED}=[shape=SEflag, minimum width=0.6cm, tikzit shape=rectangle, tikzit fill=white, fill={red!80}, draw=black, minimum height=0.6cm, line width=1, font={\footnotesize}, rotate=225, left=-0.1285cm, label={[shift={(-0.45,0.5)}]${}$}, yshift=-0.1285cm, tikzit category=mid]
\tikzstyle{REDSPIDER}=[fill={red!80}, draw=black, shape=circle, minimum size=0.2cm, line width=1, inner sep=0 cm]
\tikzstyle{BROWNSPIDER}=[fill=brown, draw=black, shape=circle, minimum size=0.2cm, line width=1, inner sep=0 cm]
\tikzstyle{OrangeSpider}=[fill=pink, draw=black, shape=circle, minimum size=0.2cm, inner sep=0 cm]
\tikzstyle{StateMidRed}=[shape=SEflag, minimum width=0.6cm, tikzit shape=rectangle, tikzit fill=red, fill={red!80}, draw=black, minimum height=0.6cm, font={\footnotesize}, rotate=45, left=-0.1285cm, label={[shift={(0.45,-0.45)}]${}$}, yshift=-0.1285cm, tikzit category=mid]
\tikzstyle{EffectMidRed}=[shape=SEflag, minimum width=0.6cm, tikzit shape=rectangle, tikzit fill=red, fill={red!80}, draw=black, minimum height=0.6cm, font={\footnotesize}, rotate=225, left=-0.1285cm, label={[shift={(-0.45,0.5)}]${}$}, yshift=-0.1285cm, tikzit category=mid]
\tikzstyle{BrownSpider}=[fill=brown, draw=black, shape=circle, minimum size=0.2cm, inner sep=0 cm]
\tikzstyle{small box}=[fill={red!80}, draw=black, shape=trapezium, minimum width=0.01cm, minimum height=0.2cm, inner sep=0.01, rotate=180, tikzit shape=rectangle, trapezium right angle=0]
\tikzstyle{small green box}=[fill=green, draw=black, shape=trapezium, minimum width=0.2cm, minimum height=0.2cm, inner sep=0.01, rotate=180, tikzit shape=rectangle, trapezium right angle=0]
\tikzstyle{small orange box}=[fill=pink, draw=black, shape=trapezium, minimum width=0.2cm, minimum height=0.2cm, inner sep=0.01, rotate=180, tikzit shape=rectangle, trapezium right angle=0]
\tikzstyle{small blue box}=[fill={blue!80}, draw=black, shape=trapezium, minimum width=0.2cm, minimum height=0.2cm, inner sep=0.01, rotate=180, tikzit shape=rectangle, trapezium right angle=0]
\tikzstyle{small brown box}=[fill=brown, draw=black, shape=trapezium, minimum width=0.2cm, minimum height=0.2cm, inner sep=0.01, rotate=180, tikzit shape=rectangle, trapezium right angle=0]
\tikzstyle{Rsmall box}=[fill={red!80}, draw=black, shape=trapezium, minimum width=0.2cm, minimum height=0.2cm, inner sep=0.01, tikzit shape=rectangle, trapezium left angle=0]
\tikzstyle{Rsmall green box}=[fill=green, draw=black, shape=trapezium, minimum width=0.2cm, minimum height=0.2cm, inner sep=0.01, tikzit shape=rectangle, trapezium left angle=0]
\tikzstyle{Rsmall orange box}=[fill=pink, draw=black, shape=trapezium, minimum width=0.2cm, minimum height=0.2cm, inner sep=0.01, tikzit shape=rectangle, trapezium left angle=0]
\tikzstyle{Rsmall blue box}=[fill={blue!80}, draw=black, shape=trapezium, minimum width=0.2cm, minimum height=0.2cm, inner sep=0.01, tikzit shape=rectangle, trapezium left angle=0]
\tikzstyle{Rsmall brown box}=[fill=brown, draw=black, shape=trapezium, minimum width=0.2cm, minimum height=0.2cm, inner sep=0.01, tikzit shape=rectangle, trapezium left angle=0]
\tikzstyle{Rflctsmall brown box}=[fill=brown, draw=black, shape=trapezium, minimum width=0.2cm, minimum height=0.2cm, inner sep=0.01, tikzit shape=rectangle, trapezium left angle=0, xscale=-1]
\tikzstyle{Rflctsmall box}=[fill={red!80}, draw=black, shape=trapezium, minimum width=0.2cm, minimum height=0.2cm, inner sep=0.01, tikzit shape=rectangle, trapezium left angle=0, xscale=-1]
\tikzstyle{Rflctsmall green box}=[fill=green, draw=black, shape=trapezium, minimum width=0.2cm, minimum height=0.2cm, inner sep=0.01, tikzit shape=rectangle, trapezium left angle=0, xscale=-1]
\tikzstyle{Rflctsmall orange box}=[fill=pink, draw=black, shape=trapezium, minimum width=0.2cm, minimum height=0.2cm, inner sep=0.01, tikzit shape=rectangle, trapezium left angle=0, xscale=-1]
\tikzstyle{Rflctsmall blue box}=[fill={blue!80}, draw=black, shape=trapezium, minimum width=0.2cm, minimum height=0.2cm, inner sep=0.01, tikzit shape=rectangle, trapezium left angle=0, xscale=-1]
\tikzstyle{small gray box}=[fill=gray, draw=black, shape=rectangle, minimum width=0.1cm, minimum height=0.1cm, inner sep=0.01, rotate=180, tikzit shape=rectangle, trapezium right angle=0]
\tikzstyle{small white box}=[fill=white, draw=black, shape=rectangle, minimum width=0.1cm, minimum height=0.1cm, inner sep=0.01, rotate=180, tikzit shape=rectangle, trapezium right angle=0]

\tikzstyle{blue dir}=[->, draw=black, dashed]
\tikzstyle{new edge style 0}=[-, dotted]
\tikzstyle{WIRE}=[-, line width=1]
\tikzstyle{DOTTEDBLUE}=[-, line width=1, dotted, draw=blue]
\tikzstyle{blue dir2}=[-, draw=black, dashed]
\tikzstyle{DOTTEDRED}=[-, dotted, line width=1, draw=red]
\tikzstyle{DOTTED}=[-, dotted, line width=1, draw=black]

\title{Diagrammatic security proof for 8-state encoding}
\author{Boris \v{S}kori\'{c} and Zef Wolffs}

\begin{document}

\date{ }
\maketitle

\setlength{\parindent}{0mm}

\begin{abstract}
\noindent
Dirac notation is the most common way to describe quantum states and operations on states. 
It is very convenient and allows for quick visual distinction between vectors, scalars and operators.
For quantum processes that involve interactions of multiple systems an even better visualisation has been proposed by Coecke
and Kissinger, in the form of a diagrammatic formalism \cite{Coecke2017}.
Their notation expresses formulas in the form of diagrams, somewhat similar to Feynman diagrams, and is more general than
the circuit notation for quantum computing.

\noindent
This document consists of two parts.
(1)
We give a brief summary of the diagrammatic notation of quantum processes,
tailored to readers who already know quantum physics and are not interested in general process theory.
For this audience our summary is less daunting than the encyclopaedic book by Coecke and Kissinger \cite{Coecke2017},
and on the other hand more accessible than the ultra-compact introduction of \cite{KTW2017}.
We deviate a somewhat from \cite{Coecke2017,KTW2017} in that
we do not assume basis states to equal their own complex conjugate; 
this means that we do not use symmetric notation for basis states, and it leads us to explicitly show arrows on wires
where they are usually omitted.

\noindent
(2)~We extend the work of Kissinger, Tull and Westerbaan \cite{KTW2017}
which gives a diagrammatic security proof for BB84 and 6-state Quantum Key Distribution.
Their proof is based on a sequence of diagrammatic
manipulations that works when the bases used in the protocol are mutually unbiased.
We extend this result to 8-state encoding, which has been proposed as a tool in
quantum key recycling protocols~\cite{SdV2017,LS2018}, and which does {\em not}
have mutually unbiased bases.
\end{abstract}


\makeatletter{\renewcommand*{\@makefnmark}{}
\footnotetext{This manuscript is based on Zef Wolffs' bachelor thesis \cite{WolffsThesis}.}\makeatother}

\section{Pure states and linear operators}
\label{section:pure}

In the diagrammatic notation of Coecke and Kissinger,
time flows from the bottom to the top, just like in Feynman diagrams.
Wires represent states. 
Boxes of various shapes represent operations.
Hermitian conjugation ($\dagger$) corresponds to vertical flipping.
Complex conjugation (c.c.) is horizontal flipping.
Taking the transpose is a 180$^\circ$ rotation.

\underline{\bf Pure states}.
The preparation of a pure state $\ket\qj$ is depicted as a triangle with a thin directed wire departing from it upwards.
The corresponding bra $\ket\qj\dagg=\bra\qj$, i.e.\;taking the inner product with $\qj$, is depicted as a thin directed
wire terminating in the vertically flipped triangle. The arrow points towards the bra.
\begin{equation}
\label{braket}
	\ket\qj \quad\equiv\quad 

\end{equation}
The picture below shows $\ket\qj$, $\bra\qj$, $\ket\qj^*$ and $\bra\qj^*$. 
Note that the wire connected to $\ket\qj^*$ and $\bra\qj^*$ has a {\em downward} arrow. 
\begin{equation}
	%
\end{equation}
Wires are sometimes labeled with the Hilbert space in which the state lives.
Basis states $\ket i$ in some orthonormal basis are often represented as a left-right symmetric shape, since their vector representation
is typically real-valued.
We will depart from this convention because we will be working with a set of bases that does not allow for such 
simultaneous real-valued representation.
We depict the decomposition $\ket\qj=\sum_i \qj_i\ket i$ as
\begin{equation}
\label{basis}
	%
\end{equation}
Note that a diagram may contain summations and scalar multiplications.
A different choice of basis is indicated as a different shading or colour of the basis triangles.
The `standard' basis is white.

\underline{\bf Inner product}.
The formula $\inprod{\qj}{\phi}^* = \inprod{\phi}{\qj}$ is depicted as
\be
	%
\label{rotinprod}
\ee
{\em Rotation of a diagram that yields a scalar has no effect on the numerical outcome.}

\underline{\bf Tensor product}.
When parts of a diagram are not connected, this indicates that the tensor product is taken.
It does not matter how these parts are arranged with respect to each other, e.g.\;vertically separated or
horizontally.
This means that {\em disconnected parts in a diagram may be freely moved around}.
\begin{equation}
	%
 \;\;\equiv\;\; \ket{\qj_1} \otimes \ket{\qj_2}
\end{equation}
Two parallel wires, one of which represents the identity on Hilbert space $\cH_1$
and the other on $\cH_2$, may be drawn as a single wire for the tensor product space
$\cH_1\otimes \cH_2$.

\underline{\bf Wires}.
A wire represents the identity map.
The diagrams below depict the spectral decomposition of the identity, 
${\mathbf 1} = \sum_i \ket i\bra i$,
and the formula 
$\inprod{\phi}{\qj}=\sum_i \inprod{\phi}{i}\inprod{i}{\qj}$ 
$=\sum_i \phi_i^*\qj_i$.
\begin{equation}
\label{equation:wiredecomposed}
	%
\end{equation}

A bent around wire has the following definition,
\begin{equation}
\label{equation:bentaroundwire}
	%
\end{equation}
This definition leads to a very handy visualisation of 
the formula $\inprod{\phi}{\qj}=\sum_i \inprod{i}{\phi}^* \inprod{i}{\qj}$
as a bent-around version of (\ref{equation:wiredecomposed}),
\begin{equation}
\label{equation:bentaroundwiresinandoutputs}
	%
\end{equation}
More generally, it was shown
in \cite{Coecke2017} that wires can be bent in any way without affecting the numerical outcome of a diagram.
This goes by the name {\em yanking equations}. For example, the S-shape bend below is equivalent to a straight wire,
\begin{equation}
\label{equation:yankingequation}
	%
\end{equation}
This is proven diagrammatically as follows.
\begin{equation}
	%
\end{equation}

\underline{\bf Linear maps}.
A linear map $A=\sum_{ij}\bra i A\ket j\; \ket i\bra j=\sum_{ij}a_{ij}\ket i\bra j$ is depicted as
\begin{equation}
	%
\end{equation}
where the input space and output space do not have to be the same.
The incoming and outgoing wire in the diagram above may stand for multiple wires.
Note that the shape of the $A$-box is asymmetric, so that vertical and horizontal flips 
are easily recognised visually.

{\em A linear map can be moved along a wire, as long as it is rotated along with the bends.}
\be
	%
\ee
The diagrammatic proof is as follows,
\be
%
\ee

\underline{\bf Spiders}.
A {\em spider} is a special linear operation from $\cH^{\otimes m}$ to $\cH^{\otimes n}$
that acts as a generalised Kronecker delta, 
\begin{equation}
\label{equation:spidermultilegdecomposed}
	%
\end{equation}
Any combination of incoming and outgoing arrows is possible.
{\em The arrows on the wires are often omitted when there is no ambiguity.}

A special case of a spider is the spider with one input and one output, which equals the identity,
\begin{equation}
\label{equation:boringspider}
	%
\end{equation}
Another special case is the spider with zero inputs and one output.
It is proportional to the equal superposition of all basis states.
\begin{equation}
\label{equation:spideroneoutput}
	%
 
\end{equation}
A spider with two outgoing arrows corresponds to the maximally entangled state $\fr1{\sqrt d}\sum_i\ket{ii}$
\be
	%
\ee
A spider that is not connected to any lines evaluates to a scalar,
namely the dimension $d$ of the Hilbert space,
\be
	%
 = \sum_i(\mbox{empty diagram})=d.
\ee
An important property of spiders is that they fuse.
\begin{equation}
\label{equation:spiderfusion}
	%
\end{equation}
This follows directly from the definition  (\ref{equation:spidermultilegdecomposed}).
A spider is defined with respect to a certain basis.
When that it not the standard basis, shading or colouring is applied, just as for basis states.

\underline{\bf Equivalence between state preparation and EPR measurement}.\\
There is a well known equivalence between on the one hand preparing a state $\ket\qj$
and on the other hand preparing a maximally entangled bipartite state 
followed by projecting one of the subsystems onto $\ket{\qj^*}$.
In diagram notation this equivalence is visualised simply by bending a wire.
\be
	\begin{tikzpicture}
		\begin{pgfonlayer}{nodelayer}
			\node [style=none] (2) at (0, 1) {};
			\node [style=none] (3) at (0, 1) {};
			\node [style=StateRight] (4) at (0, 0) {\rotatebox{90}{$\psi$}};
			\node [style=none] (5) at (0, -0.25) {};
		\end{pgfonlayer}
		\begin{pgfonlayer}{edgelayer}
			\draw (3.center) to (5.center);
			\draw[->] (0,-0.25) -- (0,0.8); 
		\end{pgfonlayer}
	\end{tikzpicture} \quad=\quad \begin{tikzpicture}
		\begin{pgfonlayer}{nodelayer}
			\node [style=none] (0) at (-2, 0) {};
			\node [style=none] (1) at (-1, -1) {};
			\node [style=none] (2) at (0, 0) {};
			\node [style=none] (6) at (-2, 1) {};
			\node [style=none] (7) at (0, 1.5) {};
			\node [style=EffectLeft] (8) at (-2, 0.8) {\rotatebox{-90}{$\psi$}};
		\end{pgfonlayer}
		\begin{pgfonlayer}{edgelayer}
			\draw [bend left=45] (2.center) to (1.center);
			\draw [bend left=45] (1.center) to (0.center);
			\draw (0.center) to (6.center);
			\draw (2.center) to (7.center);
			\draw[->] (0,0) -- (0,0.1);
			\draw[->] (-2,0.05) -- (-2,0);
		\end{pgfonlayer}
	\end{tikzpicture}
	\quad=\quad
	\begin{tikzpicture}
		\begin{pgfonlayer}{nodelayer}
			\node [style=none] (0) at (-2, 0) {};
			\node [style=none] (1) at (-1, -1) {};
			\node [style=none] (2) at (0, 0) {};
			\node [style=none] (6) at (-2, 1.5) {};
			\node [style=none] (7) at (0, 1.5) {};
			\node [style=EffectLeft] (8) at (-2, 1.5) {\rotatebox{-90}{$\psi$}};
			\node [style=WhiteSpider] (9) at (-1, -1) {};
			\node [style=WhiteSpider] (10) at (-2, 0.4) {};
		\end{pgfonlayer}
		\begin{pgfonlayer}{edgelayer}
			\draw [bend left=45] (2.center) to (1.center);
			\draw [bend left=45] (1.center) to (0.center);
			\draw (0.center) to (6.center);
			\draw (2.center) to (7.center);
			\draw[->] (0,0) -- (0,0.1);
			\draw[->] (-2,0.9) -- (-2,0.8);
			\draw[->] (-2,-0.001) -- (-2,0);
		\end{pgfonlayer}
	\end{tikzpicture}
	\quad=\quad \sum_i \qj_i  \begin{tikzpicture}
		\begin{pgfonlayer}{nodelayer}
			\node [style=none] (0) at (-2, 0) {};
			\node [style=none] (1) at (-1, -1) {};
			\node [style=none] (2) at (0, 0) {};
			\node [style=none] (6) at (-2, 1.3) {};
			\node [style=none] (7) at (0, 1.5) {};
			\node [style=EffectRight] (8) at (-2.1, 1) {\rotatebox{0}{$i$}};
			\node [style=WhiteSpider] (9) at (-1, -1) {};
		\end{pgfonlayer}
		\begin{pgfonlayer}{edgelayer}
			\draw [bend left=45] (2.center) to (1.center);
			\draw [bend left=45] (1.center) to (0.center);
			\draw (0.center) to (6.center);
			\draw (2.center) to (7.center);
			\draw[->] (0,0) -- (0,0.1);
			\draw[->] (-2,-0.001) -- (-2,0);
		\end{pgfonlayer}
	\end{tikzpicture}
	\quad=\quad   \begin{tikzpicture}
		\begin{pgfonlayer}{nodelayer}
			\node [style=none] (0) at (-2, 0) {};
			\node [style=none] (1) at (-1, -1) {};
			\node [style=none] (2) at (0, 0) {};
			\node [style=none] (6) at (-2, 1.3) {};
			\node [style=none] (7) at (0, 1.5) {};
			\node [style=EffectRight] (8) at (-2.1, 1) {{\small $\qj^*$}};
			\node [style=WhiteSpider] (9) at (-1, -1) {};
		\end{pgfonlayer}
		\begin{pgfonlayer}{edgelayer}
			\draw [bend left=45] (2.center) to (1.center);
			\draw [bend left=45] (1.center) to (0.center);
			\draw (0.center) to (6.center);
			\draw (2.center) to (7.center);
			\draw[->] (0,0) -- (0,0.1);
			\draw[->] (-2,-0.001) -- (-2,0);
		\end{pgfonlayer}
	\end{tikzpicture}
\label{EPRbend}
\ee
In the last diagram of (\ref{EPRbend}) the spider creates the entangled state, and the complex conjugation
$\qj\mapsto\qj^*$ is defined with respect to the spider's basis, i.e. $\qj_i\mapsto\qj_i^*$.

In case of a set of self-conjugate basis states the arrows on the wires do not have to be drawn,
and the spider in the last diagram in (\ref{EPRbend}) can be omitted, resulting in a simple 
picture consisting merely of the $\bra{\qj^*}$ bra and a bent wire.

\underline{\bf Phase spiders}.
Let $\qa=(\qa_i)_{i=0}^{d-1}$, with $\qa_0=0$ by convention, be a vector containing angles.
The $\qa$-phase spider is defined as
\be
	\begin{tikzpicture}
		\begin{pgfonlayer}{nodelayer}
			\node [style=WhiteSpider] (0) at (0, 0) {$\alpha$};
			\node [style=none] (1) at (-1, 1) {};
			\node [style=none] (2) at (-0.5, 1) {};
			\node [style=none] (3) at (1, 1) {};
			\node [style=none] (4) at (-1, -1.25) {};
			\node [style=none] (5) at (-0.5, -1.25) {};
			\node [style=none] (6) at (1, -1.25) {};
			\node [style=none] (7) at (-1, 1) {};
			\node [style=none] (8) at (0, -1.25) {};
			\node [style=none] (9) at (0.5, -1.25) {};
			\node [style=none] (10) at (0, 1) {};
			\node [style=none] (11) at (0.5, 1) {};
			\node [style=none] (12) at (0, -1.75) {\small $m$ inputs};
			\node [style=none] (13) at (0, 1.5) {\small $n$ outputs};
		\end{pgfonlayer}
		\begin{pgfonlayer}{edgelayer}
			\draw [bend left] (0) to (7.center);
			\draw [bend left=15] (0) to (2.center);
			\draw [bend right] (0) to (3.center);
			\draw [bend right] (0) to (4.center);
			\draw [bend right=15] (0) to (5.center);
			\draw [bend left] (0) to (6.center);
			\draw [style=new edge style 0] (10.center) to (11.center);
			\draw [style=new edge style 0] (8.center) to (9.center);
		\end{pgfonlayer}
	\end{tikzpicture}
	\quad\isdef\quad
	\sum_j e^{i\qa_j}\;  \begin{tikzpicture}
		\begin{pgfonlayer}{nodelayer}
			\node [style=none] (7) at (-3, 2.5) {};
			\node [style=none] (8) at (-3, -2.5) {};
			\node [style=none] (9) at (-3, 1) {};
			\node [style=none] (10) at (-3, -1) {};
			\node [style=none] (14) at (-1, 2.5) {};
			\node [style=none] (15) at (-1, -2.5) {};
			\node [style=none] (16) at (-1, 1) {};
			\node [style=none] (17) at (-1, -1) {};
			\node [style=none] (26) at (2, 2.5) {};
			\node [style=none] (27) at (2, -2.5) {};
			\node [style=none] (28) at (2, 1) {};
			\node [style=none] (29) at (2, -1) {};
			\node [style=none] (32) at (0, 2) {};
			\node [style=none] (33) at (1, 2) {};
			\node [style=none] (34) at (0, -2) {};
			\node [style=none] (35) at (1, -2) {};
			\node [style=none] (36) at (-0.5, 3) {\small $n$ outputs};
			\node [style=none] (37) at (-0.5, -3) {\small $m$ inputs};
			\node [style=StateRight] (38) at (-3, 1) {\rotatebox{90}{j}};
			\node [style=EffectRight] (39) at (-3, -1) {\rotatebox{0}{j}};
			\node [style=StateRight] (40) at (-1, 1) {\rotatebox{90}{j}};
			\node [style=StateRight] (41) at (2, 1) {\rotatebox{90}{j}};
			\node [style=EffectRight] (42) at (-1, -1) {\rotatebox{0}{j}};
			\node [style=EffectRight] (43) at (2, -1) {\rotatebox{0}{j}};
		\end{pgfonlayer}
		\begin{pgfonlayer}{edgelayer}
			\draw (7.center) to (9.center);
			\draw (8.center) to (10.center);
			\draw (14.center) to (16.center);
			\draw (15.center) to (17.center);
			\draw (26.center) to (28.center);
			\draw (27.center) to (29.center);
			\draw [style=new edge style 0] (32.center) to (33.center);
			\draw [style=new edge style 0] (34.center) to (35.center);
		\end{pgfonlayer}
	\end{tikzpicture}
\label{phasespider}
\ee 
{\em Phase spiders fuse like normal spiders, but their phases are added.}
This follows directly from definitions (\ref{equation:spidermultilegdecomposed}) and (\ref{phasespider}).
\be
	\begin{tikzpicture}
		\begin{pgfonlayer}{nodelayer}
			\node [style=WhiteSpider] (0) at (-0.75, 0.5) {\small $\alpha$};
			\node [style=none] (1) at (-1.75, 1.5) {};
			\node [style=none] (2) at (-1.25, 1.5) {};
			\node [style=none] (3) at (0.25, 1.5) {};
			\node [style=none] (4) at (-2, -1.75) {};
			\node [style=none] (5) at (-1.5, -1.75) {};
			\node [style=none] (6) at (0, 0) {};
			\node [style=none] (7) at (-1.75, 1.5) {};
			\node [style=none] (8) at (-1, -1.75) {};
			\node [style=none] (9) at (-0.5, -1.75) {};
			\node [style=none] (10) at (-0.75, 1.5) {};
			\node [style=none] (11) at (-0.25, 1.5) {};
			\node [style=WhiteSpider] (12) at (1, -0.5) {\small $\beta$};
			\node [style=none] (13) at (0, 0) {};
			\node [style=none] (14) at (0.75, 1.5) {};
			\node [style=none] (15) at (2.25, 1.5) {};
			\node [style=none] (16) at (0, -1.75) {};
			\node [style=none] (17) at (0.5, -1.75) {};
			\node [style=none] (18) at (2, -1.75) {};
			\node [style=none] (19) at (0, 0) {};
			\node [style=none] (20) at (1, -1.75) {};
			\node [style=none] (21) at (1.5, -1.75) {};
			\node [style=none] (22) at (1.25, 1.5) {};
			\node [style=none] (23) at (1.75, 1.5) {};
			\node [style=none] (24) at (0.25, 2) {\small $n$ outputs};
			\node [style=none] (25) at (0.25, -2.25) {\small $m$ inputs};
		\end{pgfonlayer}
		\begin{pgfonlayer}{edgelayer}
			\draw [bend left] (0) to (7.center);
			\draw [bend left=15] (0) to (2.center);
			\draw [bend right] (0) to (3.center);
			\draw [bend right] (0) to (4.center);
			\draw [bend right=15] (0) to (5.center);
			\draw [bend left] (0) to (6.center);
			\draw [style=new edge style 0] (10.center) to (11.center);
			\draw [style=new edge style 0] (8.center) to (9.center);
			\draw [bend left] (12) to (19.center);
			\draw [bend left=15] (12) to (14.center);
			\draw [bend right] (12) to (15.center);
			\draw [bend right] (12) to (16.center);
			\draw [bend right=15] (12) to (17.center);
			\draw [bend left] (12) to (18.center);
			\draw [style=new edge style 0] (22.center) to (23.center);
			\draw [style=new edge style 0] (20.center) to (21.center);
		\end{pgfonlayer}
	\end{tikzpicture} = \begin{tikzpicture}
	\begin{pgfonlayer}{nodelayer}
		\node [style=WhiteSpider] (0) at (0, 0) {\small $\alpha{+}\beta$};
		\node [style=none] (1) at (-1, 1) {};
		\node [style=none] (2) at (-0.5, 1) {};
		\node [style=none] (3) at (1, 1) {};
		\node [style=none] (4) at (-1, -1.25) {};
		\node [style=none] (5) at (-0.5, -1.25) {};
		\node [style=none] (6) at (1, -1.25) {};
		\node [style=none] (7) at (-1, 1) {};
		\node [style=none] (8) at (0, -1.25) {};
		\node [style=none] (9) at (0.5, -1.25) {};
		\node [style=none] (10) at (0, 1) {};
		\node [style=none] (11) at (0.5, 1) {};
		\node [style=none] (12) at (0, -1.75) {\small $m$ inputs};
		\node [style=none] (13) at (0, 1.5) {\small $n$ outputs};
	\end{pgfonlayer}
	\begin{pgfonlayer}{edgelayer}
		\draw [bend left] (0) to (7.center);
		\draw [bend left=15] (0) to (2.center);
		\draw [bend right] (0) to (3.center);
		\draw [bend right] (0) to (4.center);
		\draw [bend right=15] (0) to (5.center);
		\draw [bend left] (0) to (6.center);
		\draw [style=new edge style 0] (10.center) to (11.center);
		\draw [style=new edge style 0] (8.center) to (9.center);
	\end{pgfonlayer}
\end{tikzpicture}
\ee
The complex conjugate of a phase spider has flipped phases.
\be
	\begin{tikzpicture}
		\begin{pgfonlayer}{nodelayer}
			\node [style=WhiteSpider] (0) at (0, 0) {\,$\alpha$\,};
		\end{pgfonlayer}
	\end{tikzpicture}^* = \begin{tikzpicture}
	\begin{pgfonlayer}{nodelayer}
		\node [style=WhiteSpider] (0) at (0, 0) {$-\alpha$};
	\end{pgfonlayer}
\end{tikzpicture}
\label{equation:conjugatespider}
\ee

\underline{\bf Qubit states; Bloch sphere; switchable Pauli operations}.

Qubit states ($d=2$) correspond to spin states on the Bloch sphere.
Typically the $z$-basis is chosen as the standard basis, with $\ket 0=\ket{+z}$ and
$\ket 1=\ket{-z}$.
The Pauli spin matrices are $\qs_x={0\; 1\choose 1\;0}$,
$\qs_y={0 \; -i \choose i\; \phantom{-}0}$,
$\qs_z={1\; \phantom{-}0 \choose 0\; -1}$.
A general spin state with spherical angles $(\qy,\qf)$ is given by $\cos\fr\qy2\ket0+ e^{i\qf}\sin\fr\qy2\ket1$.
The spin states along the $x,y$ axes are given by
$\ket{\pm x}=\frac{\ket0\pm\ket 1}{\sqrt2}$, 
$\ket{\pm y}=\frac{\ket0\pm i \ket 1}{\sqrt2}$.
The `$+$' state in each case represents a logical~$0$.

In many qubit-based protocols it suffices to work with only the $z$-basis and the $x$-basis.
The convention in \cite{Coecke2017,KTW2017} is to use white colour for the $z$-basis and
grey shading for the $x$-basis. The basis states are depicted as left-right symmetric, since they are real-valued 
when expressed in the standard basis.
The following diagrammatic relations hold between the two bases,
\be
	\begin{tikzpicture}
		\begin{pgfonlayer}{nodelayer}
			\node [style=WhiteSpider] (0) at (0, -0.5) {};
			\node [style=none] (1) at (0, 0.5) {};
		\end{pgfonlayer}
		\begin{pgfonlayer}{edgelayer}
			\draw (0) to (1.center);
		\end{pgfonlayer}
	\end{tikzpicture} = \sqrt2\, \begin{tikzpicture}
	\begin{pgfonlayer}{nodelayer}
		\node [style=StateMidGray] (0) at (0, 0) {\rotatebox{135}{0}};
		\node [style=none] (1) at (0, -0.5) {};
		\node [style=none] (2) at (0, 1) {};
	\end{pgfonlayer}
	\begin{pgfonlayer}{edgelayer}
		\draw [in=270, out=90] (1.center) to (2.center);
	\end{pgfonlayer}
\end{tikzpicture}
	\quad\quad
	\begin{tikzpicture}
		\begin{pgfonlayer}{nodelayer}
			\node [style=WhiteSpider] (0) at (0, -0.5) {$\pi$};
			\node [style=none] (1) at (0, 0.5) {};
		\end{pgfonlayer}
		\begin{pgfonlayer}{edgelayer}
			\draw (0) to (1.center);
		\end{pgfonlayer}
	\end{tikzpicture} = \sqrt2\, \begin{tikzpicture}
	\begin{pgfonlayer}{nodelayer}
		\node [style=StateMidGray] (0) at (0, 0) {\rotatebox{-45}{1}};
		\node [style=none] (1) at (0, -0.5) {};
		\node [style=none] (2) at (0, 1) {};
	\end{pgfonlayer}
	\begin{pgfonlayer}{edgelayer}
		\draw [in=270, out=90] (1.center) to (2.center);
	\end{pgfonlayer}
\end{tikzpicture}
	\quad\quad
	\begin{tikzpicture}
		\begin{pgfonlayer}{nodelayer}
			\node [style=GraySpider] (0) at (0, -0.5) {};
			\node [style=none] (1) at (0, 0.5) {};
		\end{pgfonlayer}
		\begin{pgfonlayer}{edgelayer}
			\draw (0) to (1.center);
		\end{pgfonlayer}
	\end{tikzpicture} = \sqrt2\, \begin{tikzpicture}
	\begin{pgfonlayer}{nodelayer}
		\node [style=StateMid] (0) at (0, 0) {\rotatebox{135}{0}};
		\node [style=none] (1) at (0, 1) {};
		\node [style=none] (2) at (0, -0.5) {};
	\end{pgfonlayer}
	\begin{pgfonlayer}{edgelayer}
		\draw (2.center) to (1.center);
	\end{pgfonlayer}
\end{tikzpicture}
	\quad\quad
	\begin{tikzpicture}
		\begin{pgfonlayer}{nodelayer}
			\node [style=GraySpider] (0) at (0, -0.5) {$\pi$};
			\node [style=none] (1) at (0, 0.5) {};
		\end{pgfonlayer}
		\begin{pgfonlayer}{edgelayer}
			\draw (0) to (1.center);
		\end{pgfonlayer}
	\end{tikzpicture} = \sqrt2\, \begin{tikzpicture}
	\begin{pgfonlayer}{nodelayer}
		\node [style=StateMid] (0) at (0, 0) {\rotatebox{-45}{1}};
		\node [style=none] (1) at (0, 1) {};
		\node [style=none] (2) at (0, -0.5) {};
	\end{pgfonlayer}
	\begin{pgfonlayer}{edgelayer}
		\draw (2.center) to (1.center);
	\end{pgfonlayer}
\end{tikzpicture}
\ee
Here a phase label $\pi$ is shorthand for the two-dimensional phase vector $(0,\pi)$.
Furthermore,
a white $\pi$-spider with a single in- and output acts on a state like the Pauli operation $\qs_z$,
and a gray $\pi$-spider acts like $\qs_x$.
\be
	\begin{tikzpicture}
		\begin{pgfonlayer}{nodelayer}
			\node [style=WhiteSpider] (0) at (0, 0) {$\pi$};
			\node [style=none] (1) at (0, 1.0) {};
			\node [style=none] (2) at (0, -1.0) {};
		\end{pgfonlayer}
		\begin{pgfonlayer}{edgelayer}
			\draw (0) to (1.center);
			\draw (2) to (1.center);
		\end{pgfonlayer}
	\end{tikzpicture} = \begin{tikzpicture}
	\begin{pgfonlayer}{nodelayer}
		\node [style=medium box] (0) at (0, 0) {$\sigma_z$};
		\node [style=none] (1) at (0, 1.0) {};
		\node [style=none] (2) at (0, -1.0) {};
	\end{pgfonlayer}
	\begin{pgfonlayer}{edgelayer}
		\draw (0) to (1.center);
		\draw (2) to (1.center);
	\end{pgfonlayer}
\end{tikzpicture}
	\quad\quad\quad\quad
	\begin{tikzpicture}
		\begin{pgfonlayer}{nodelayer}
			\node [style=GraySpider] (0) at (0, 0) {$\pi$};
			\node [style=none] (1) at (0, 1.0) {};
			\node [style=none] (2) at (0, -1.0) {};
		\end{pgfonlayer}
		\begin{pgfonlayer}{edgelayer}
			\draw (0) to (1.center);
			\draw (2) to (1.center);
		\end{pgfonlayer}
	\end{tikzpicture} = \begin{tikzpicture}
	\begin{pgfonlayer}{nodelayer}
		\node [style=medium box] (0) at (0, 0) {$\sigma_x$};
		\node [style=none] (1) at (0, 1.0) {};
		\node [style=none] (2) at (0, -1.0) {};
	\end{pgfonlayer}
	\begin{pgfonlayer}{edgelayer}
		\draw (0) to (1.center);
		\draw (2) to (1.center);
	\end{pgfonlayer}
\end{tikzpicture}
\ee
Switchable $\qs_z$ and $\qs_x$ operations are built as follows.
\be
	\sqrt2\begin{tikzpicture}
		\begin{pgfonlayer}{nodelayer}
			\node [style=WhiteSpider] (0) at (0, 0) {};
			\node [style=none] (1) at (0, 1.0) {};
			\node [style=none] (2) at (-1, -1.0) {};
			\node [style=StateMidGray] (3) at (1, -1) {\rotatebox{-45}{u}};
		\end{pgfonlayer}
		\begin{pgfonlayer}{edgelayer}
			\draw (1) to (0.center);
			\draw (2) to (0.center);
			\draw (3.center) to (0.center);
		\end{pgfonlayer}
	\end{tikzpicture} = \begin{tikzpicture}
	\begin{pgfonlayer}{nodelayer}
		\node [style=WhiteSpider] (0) at (0, 0) {$u\pi$};
		\node [style=none] (1) at (0, 1.0) {};
		\node [style=none] (2) at (0, -1.0) {};
	\end{pgfonlayer}
	\begin{pgfonlayer}{edgelayer}
		\draw (0) to (1.center);
		\draw (2) to (1.center);
	\end{pgfonlayer}
\end{tikzpicture} =\begin{tikzpicture}
\begin{pgfonlayer}{nodelayer}
	\node [style=medium box] (0) at (0, 0) {$\sigma_z^u$};
	\node [style=none] (1) at (0, 1.0) {};
	\node [style=none] (2) at (0, -1.0) {};
\end{pgfonlayer}
\begin{pgfonlayer}{edgelayer}
	\draw (0) to (1.center);
	\draw (2) to (1.center);
\end{pgfonlayer}
\end{tikzpicture}
	\quad\quad\quad\quad\quad\quad
	\sqrt2\begin{tikzpicture}
		\begin{pgfonlayer}{nodelayer}
			\node [style=GraySpider] (0) at (0, 0) { };
			\node [style=none] (1) at (0, 1.0) {};
			\node [style=none] (2) at (-1, -1.0) {};
			\node [style=StateMid] (3) at (1, -1) {\rotatebox{-45}{u}};
		\end{pgfonlayer}
		\begin{pgfonlayer}{edgelayer}
			\draw (1) to (0.center);
			\draw (2) to (0.center);
			\draw (3.center) to (0.center);
		\end{pgfonlayer}
	\end{tikzpicture} =\begin{tikzpicture}
	\begin{pgfonlayer}{nodelayer}
		\node [style=GraySpider] (0) at (0, 0) {$u\pi$};
		\node [style=none] (1) at (0, 1.0) {};
		\node [style=none] (2) at (0, -1.0) {};
	\end{pgfonlayer}
	\begin{pgfonlayer}{edgelayer}
		\draw (0) to (1.center);
		\draw (2) to (1.center);
	\end{pgfonlayer}
\end{tikzpicture} =\begin{tikzpicture}
\begin{pgfonlayer}{nodelayer}
	\node [style=medium box] (0) at (0, 0) {$\sigma_x^u$};
	\node [style=none] (1) at (0, 1.0) {};
	\node [style=none] (2) at (0, -1.0) {};
\end{pgfonlayer}
\begin{pgfonlayer}{edgelayer}
	\draw (0) to (1.center);
	\draw (2) to (1.center);
\end{pgfonlayer}
\end{tikzpicture}
\label{encspiders}
\ee
Here the $u\in\bits$ is a control bit that switches the Pauli operation. 

{\bf Quantum One-Time Pad} (QOTP) encryption \cite{AMTdW2000,Leung2002,BR2003} of a qubit state 
works with a two-bit key $(u,v)\in\bits^2$. 
The encryption operation $E$ is
\be
	E_{uv}\ket\qj \isdef \qs_z^u \qs_x^v \ket\qj
\ee
which can be depicted as
\be
	E_{uv}\ket\qj  \quad\equiv\quad 2\begin{tikzpicture}
		\begin{pgfonlayer}{nodelayer}
			\node [style=StateRight] (0) at (0, -1) {\rotatebox{90}{$\psi$}};
			\node [style=GraySpider] (1) at (0, 0) {};
			\node [style=WhiteSpider] (2) at (0, 1.5) {};
			\node [style=StateMid] (3) at (2, -0.5) {\rotatebox{-45}{v}};
			\node [style=StateMidGray] (4) at (2, 1) {\rotatebox{-45}{u}};
			\node [style=none] (10) at (0, 2) {};
		\end{pgfonlayer}
		\begin{pgfonlayer}{edgelayer}
			\draw (0.center) to (10);
			\draw [out=90, in=0] (3.center) to (1.center);
			\draw [out=90, in=0] (4.center) to (2.center);
		\end{pgfonlayer}
	\end{tikzpicture}
\label{QOTPthin}
\ee
\underline{\bf Bell states}.
The Bell states in two-qubit space are usually written as 
$\ket{\qF^\pm}=\frac{\ket{00}\pm\ket{11}}{\sqrt2}$,
$\ket{\qJ^\pm}=\frac{\ket{01}\pm\ket{10}}{\sqrt2}$.
We write them in a slightly different way which
displays the fact that the four Bell states are QOTP encryptions of each other, where the encryption acts on one of the two qubits,
\be
	\ket{\qF_{uv}} = ({\bf 1}\otimes\qs_z^u \qs_x^v) \frac{\ket{00}+\ket{11}}{\sqrt2}
	=\frac{\ket{0v}+(-1)^u\ket{1\bar v}}{\sqrt2}.
\ee
\be
	\begin{tikzpicture}
		\begin{pgfonlayer}{nodelayer}
			\node [style=none] (0) at (0, -1) {};
			\node [style=GraySpider] (1) at (0, 0) {};
			\node [style=WhiteSpider] (2) at (0, 1.5) {};
			\node [style=StateMid] (3) at (2, -0.5) {\rotatebox{-45}{v}};
			\node [style=StateMidGray] (4) at (2, 1) {\rotatebox{-45}{u}};
			\node [style=none] (10) at (0, 2) {};
			\node [style=WhiteSpider] (20) at (-0.5, -1.5) {};
			\node [style=none] (21) at (-1, -1) {};
			\node [style=none] (22) at (-1, 2) {};
		\end{pgfonlayer}
		\begin{pgfonlayer}{edgelayer}
			\draw (0.center) to (10);
			\draw [out=90, in=0] (3.center) to (1.center);
			\draw [out=90, in=0] (4.center) to (2.center);
			\draw [out=-90, in=0]   (0.center) to (20.center);
			\draw [out=180, in=-90] (20.center) to (21.center);
			\draw (21.center) to (22.center);
			\draw[->] (-1,-1.1) -- (-1,-1);
			\draw[->] (0,-1.1) -- (0,-1);
		\end{pgfonlayer}
	\end{tikzpicture} \quad\equiv\quad \sqrt2 \ket{\qF_{uv}}.
\label{BellStateuv}
\ee

\section{Mixed states}
\label{doubling}

\underline{\bf Doubling}.
In the standard formalism, mixed states are described as positive semidefinite operators on
the Hilbert space $\cH$, with trace~1.
The space of such {\em density matrices} is denoted as $\cD(\cH)$.
Because of the isomorphism
\be
	\cD(\cH)\cong \cH\otimes\cH
\ee
it is possible to express a mixed state, which is usually written in operator form
\footnote{
We will use tilde notation for the operator form.
} as $\tilde\qr=\sum_{ij}\qr_{ij}\ket i\bra j\in\cD(\cH)$,
as an object in $\cH\otimes\cH$, namely $\qr= \sum_{ij}\qr_{ij}\ket i^*\otimes\ket j$.
This has the advantage that it becomes possible to draw diagrams for mixed states in a way that is visually similar to pure states,
i.e.\;creation of a mixed state has wires extending only upward, whereas an operator box has wires up and down.
For a pure state $\ket\qj=\sum_i \qj_i \ket i$, whose density matrix is $\ket\qj\bra\qj\in\cD(\cH)$,
this yields
$\qr=\sum_{ij}\qj_i^* \qj_j \ket i^*\otimes \ket j$
$=\ket\qj^*\otimes\ket\qj$, with diagram
\be
	\begin{tikzpicture}
		\begin{pgfonlayer}{nodelayer}
			\node [style=StateLeft] (0) at (-0.5, -0.6) {\rotatebox{180}{$\psi$}};
			\node [style=StateRight] (1) at (0.5, -0.5) {\;\rotatebox{90}{$\psi$}};
			\node [style=none] (2) at (-0.5, 1) {};
			\node [style=none] (3) at (0.5, 1) {};
		\end{pgfonlayer}
		\begin{pgfonlayer}{edgelayer}
			\draw (2) to (0.center);
			\draw (3) to (1.center);
			\draw[->] (0.5,0) -- (0.5,0.3);
			\draw[->] (-0.5,0.5) -- (-0.5,0.2);
		\end{pgfonlayer}
	\end{tikzpicture}
\label{psistarpsi}
\ee
For brevity a special notation is introduced, in which a {\bf thick-drawn} wire, spider, pure state 
or pure map\footnote{
A pure map is an operation on mixed states $\tilde\qr\in\cD(\cH)$
that can be represented as $A\tilde\qr A\dagg$, i.e.~a linear operation on~$\cH$. 
}
indicates {\bf doubling} of the object
as in (\ref{psistarpsi}), adding a complex conjugated copy to the left.
Any thick-lined box/triangle means that it pertains to {\em mixed} states only.
It is not possible to have a thick box or spider connected to a thin wire.
The special case of pure states and pure operations
is indicated with a `hat' notation,
\be
\label{equation:doubledstate}
	\begin{tikzpicture}
		\begin{pgfonlayer}{nodelayer}
			\node [style=STATERIGHT] (0) at (0, -0.5) {\rotatebox{90}{\scriptsize $\hat{\psi} \vphantom{M^{M^2}}$}};
			\node [style=none] (1) at (0, -1) {};
			\node [style=none] (2) at (0, 1) {};
			\node [style=none] (3) at (-0.7, 0.5) {{\tiny$\cD(\mathcal{H})$}};
		\end{pgfonlayer}
		\begin{pgfonlayer}{edgelayer}
			\draw [style=WIRE] (1.center) to (2.center);
		\end{pgfonlayer}
	\end{tikzpicture} \quad\isdef\quad \begin{tikzpicture}
	\begin{pgfonlayer}{nodelayer}
		\node [style=StateLeft] (0) at (-0.5, -0.6) {\rotatebox{180}{$\psi$}};
		\node [style=StateRight] (1) at (0.5, -0.5) {\;\rotatebox{90}{$\psi$}};
		\node [style=none] (2) at (-0.5, 1) {};
		\node [style=none] (3) at (0.5, 1) {};
		\node [style=none] (10) at (0.8, 0.7) {{\tiny$\mathcal{H}$}};
		\node [style=none] (11) at (-0.8, 0.7) {{\tiny$\mathcal{H}$}};
	\end{pgfonlayer}
	\begin{pgfonlayer}{edgelayer}
		\draw (2) to (0.center);
		\draw (3) to (1.center);
		\draw[->] (0.5,0) -- (0.5,0.3);
		\draw[->] (-0.5,0.5) -- (-0.5,0.2);
	\end{pgfonlayer}
\end{tikzpicture}
	\quad\quad\quad\quad\quad\quad
	\begin{tikzpicture}
		\begin{pgfonlayer}{nodelayer}
			\node [style=ASYM BOX] (0) at (0, 0) {\reflectbox{$\hat{A}$}};
			\node [style=none] (1) at (0.25, 1.5) {};
			\node [style=none] (2) at (0.25, -1.5) {};
			\node [style=none] (4) at (1, 1.25) {\tiny$\cD({\mathcal H}_{\!4})$};
			\node [style=none] (5) at (1, -1.25) {\tiny$\cD({\mathcal H}_{\!2})$};
			\node [style=none] (6) at (-0.75, 1.5) {};
			\node [style=none] (7) at (-0.75, -1.5) {};
			\node [style=none] (8) at (-1.5, 1.25) {\tiny$\cD({\mathcal H}_{\!3})$};
			\node [style=none] (9) at (-1.5, -1.25) {\tiny${\cD(\mathcal H}_{\!1})$};
		\end{pgfonlayer}
		\begin{pgfonlayer}{edgelayer}
			\draw [style=WIRE] (1.center) to (2.center);
			\draw [style=WIRE] (6.center) to (7.center);
		\end{pgfonlayer}
	\end{tikzpicture} \quad\isdef\quad \begin{tikzpicture}
	\begin{pgfonlayer}{nodelayer}
		\node [style=asym box] (1) at (1.25, 0) {\reflectbox{$A$}};
		\node [style=asym box conjugate] (2) at (-1.5, 0) {$A$};
		\node [style=none] (3) at (0.75, 0) {};
		\node [style=none] (4) at (-0.5, 0) {};
		\node [style=none] (13) at (-2.75, -2.25) {\tiny$\mathcal{H}_1$};
		\node [style=none] (14) at (-1, -2.25) {\tiny$\mathcal{H}_1$};
		\node [style=none] (15) at (-2.75, 2.25) {\tiny$\mathcal{H}_3$};
		\node [style=none] (16) at (-1, 2.25) {\tiny$\mathcal{H}_3$};
		\node [style=none] (17) at (-1.5, 2.25) {};
		\node [style=none] (18) at (-2.25, 2.25) {};
		\node [style=none] (19) at (-2.25, -2.25) {};
		\node [style=none] (20) at (-1.5, -2.25) {};
		\node [style=none] (21) at (0, 1.25) {};
		\node [style=none] (22) at (-1.75, 1.25) {};
		\node [style=none] (23) at (0.75, 0) {};
		\node [style=none] (24) at (-0.5, 0) {};
		\node [style=none] (25) at (-1.5, -2.25) {};
		\node [style=none] (26) at (-2.25, -2.25) {};
		\node [style=none] (27) at (0, -1.25) {};
		\node [style=none] (28) at (-1.75, -1.25) {};
		\node [style=none] (29) at (2, 2.25) {};
		\node [style=none] (30) at (2.75, 2.25) {};
		\node [style=none] (31) at (0, 1.1) {};
		\node [style=none] (32) at (2.25, 1.25) {};
		\node [style=none] (33) at (-2, 0) {};
		\node [style=none] (34) at (1.25, 0) {};
		\node [style=none] (35) at (2, -2.25) {};
		\node [style=none] (36) at (2.75, -2.25) {};
		\node [style=none] (37) at (0, -1.1) {};
		\node [style=none] (38) at (2.25, -1.25) {};
		\node [style=none] (39) at (-2, 0) {};
		\node [style=none] (40) at (1.25, 0) {};
		\node [style=none] (41) at (1.5, 2.25) {\tiny$\mathcal{H}_4$};
		\node [style=none] (42) at (3.25, 2.25) {\tiny$\mathcal{H}_4$};
		\node [style=none] (43) at (1.5, -2.25) {\tiny$\mathcal{H}_2$};
		\node [style=none] (44) at (3.25, -2.25) {\tiny$\mathcal{H}_2$};
	\end{pgfonlayer}
	\begin{pgfonlayer}{edgelayer}
		\draw [bend right=45, looseness=0.50] (4.center) to (22.center);
		\draw [bend left] (22.center) to (18.center);
		\draw [bend left, looseness=1.25] (21.center) to (17.center);
		\draw [bend right=45, looseness=0.75] (3.center) to (21.center);
		\draw [bend left=45, looseness=0.50] (24.center) to (28.center);
		\draw [bend right] (28.center) to (26.center);
		\draw [bend right, looseness=1.25] (27.center) to (25.center);
		\draw [bend left=45, looseness=0.75] (23.center) to (27.center);
		\draw [bend right] (32.center) to (30.center);
		\draw [bend right, looseness=1.25] (31.center) to (29.center);
		\draw [bend right, looseness=1.25] (31.center) to (33.center);
		\draw [bend right, looseness=0.75] (32.center) to (34.center);
		\draw [bend left] (38.center) to (36.center);
		\draw [bend left, looseness=1.25] (37.center) to (35.center);
		\draw [bend left, looseness=1.25] (37.center) to (39.center);
		\draw [bend left, looseness=0.75] (38.center) to (40.center);
	\end{pgfonlayer}
\end{tikzpicture}
\ee
Thick spiders fuse just like thin spiders.
Non-pure states $\qr$ and non-pure operators $A$ lack the left-right separated form,
but can always be written in decomposed form, e.g. 
\be
	 \begin{tikzpicture}
	 	\begin{pgfonlayer}{nodelayer}
	 		\node [style=STATERIGHT] (0) at (0, -0.625) {\rotatebox{90}{$\rho$}};
	 		\node [style=none] (1) at (0, 0.625) {};
	 		\node [style=none] (2) at (0, -1.125) {};
	 	\end{pgfonlayer}
	 	\begin{pgfonlayer}{edgelayer}
	 		\draw [style=WIRE] (1.center) to (2.center);
	 	\end{pgfonlayer}
	 \end{tikzpicture} \quad\isdef\quad
	 \sum_{ij} \rho_{ij} \begin{tikzpicture}
	 	\begin{pgfonlayer}{nodelayer}
	 		\node [style=StateLeft] (0) at (-0.5, -0.6) {\rotatebox{180}{$i$}};
	 		\node [style=StateRight] (1) at (0.5, -0.5) {\;\rotatebox{90}{$j$}};
	 		\node [style=none] (2) at (-0.5, 1) {};
	 		\node [style=none] (3) at (0.5, 1) {};
	 	\end{pgfonlayer}
	 	\begin{pgfonlayer}{edgelayer}
	 		\draw (2) to (0.center);
	 		\draw (3) to (1.center);
	 		\draw[->] (0.5,0) -- (0.5,0.3);
	 		\draw[->] (-0.5,0.5) -- (-0.5,0.2);
	 	\end{pgfonlayer}
	 \end{tikzpicture}
\label{rhoexpand}
\ee
While it is not possible for a thick spider to connect to a thin wire, the reverse is possible
(thin spider, thick wire)
since two parallel identical thin wires may be interpreted as a thick one.
The term `bastard spider' is used for a thin spider connected to at least one thick wire.
A number of important examples are shown below,
\be
	\begin{tikzpicture}
		\begin{pgfonlayer}{nodelayer}
			\node [style=WhiteSpider] (0) at (0, -1) {};
			\node [style=none] (1) at (0, 1) {};
		\end{pgfonlayer}
		\begin{pgfonlayer}{edgelayer}
			\draw [style=WIRE] (1) to (0.center);
		\end{pgfonlayer}
	\end{tikzpicture}= d\cdot\{\mbox{fully mixed state}\}
	\quad\quad\quad\quad\quad
	\begin{tikzpicture}
		\begin{pgfonlayer}{nodelayer}
			\node [style=WhiteSpider] (1) at (0, 0) {};
			\node [style=none] (2) at (0, -1.25) {};
			\node [style=none] (3) at (0, 1.25) {};
		\end{pgfonlayer}
		\begin{pgfonlayer}{edgelayer}
			\draw (1) to (2.center);
			\draw [style=WIRE] (3.center) to (1);
		\end{pgfonlayer}
	\end{tikzpicture} \mbox{\;``encoding''}
	\quad\quad\quad\quad\quad
	\begin{tikzpicture}
		\begin{pgfonlayer}{nodelayer}
			\node [style=WhiteSpider] (1) at (0, 0) {};
			\node [style=none] (2) at (0, -1.25) {};
			\node [style=none] (3) at (0, 1.25) {};
		\end{pgfonlayer}
		\begin{pgfonlayer}{edgelayer}
			\draw [style=WIRE] (1) to (2.center);
			\draw (3.center) to (1);
		\end{pgfonlayer}
	\end{tikzpicture} \mbox{\;``decoding''}
\label{spidermixed}
\ee
The `encoding' operation creates the diagonal matrix element $\ket{ii}$ from the basis state $\ket i$;
`decoding' does the opposite.
These operations allow for a description of quantum-classical systems such that thick wires represent quantum states
and thin wires stand for classical states. (A classical value $i\in\{0,\ldots,d-1\}$ is transported as 
the basis state $\ket i$ on a thin wire).
Encoding and decoding in the same basis acts as the identity.
Decoding in a basis {\em complementary} to the encoding (orthogonal on the Bloch sphere)
leads to a failure to convey data, which is diagrammatically visible as a {\bf disconnect},
\be

\label{equation:ComplementarityPositive}
\ee

The fusion of a thin spider and a thick spider leads to a bastard spider.
\begin{equation}
\label{equation:bastardspiderfusion}
	%
\end{equation}

\underline{\bf Taking the trace}.
The (partial) trace of a state is taken by connecting two whin wires that together make up a thick wire.
This operation is depicted with the `ground' symbol.
\be
\label{equation:discarding}
	%
\ee
Another name for this operation is `discarding', since the thick wire terminates.
In the operator formulation of mixed states it holds that $\tr\tilde\qr=1$.
Diagrammatically this is expressed as
\be
\label{equation:discardingstate}
	%
	\equaltext{(\ref{equation:bentaroundwiresinandoutputs})} \sum_{ij} \rho_{ij} \qd_{ij} 
	= \sum\limits_{i} \rho_{ii} =1.
\ee
For a bipartite system,
taking the partial trace $\tr_{\!\rm A}\tilde\qr^{\rm AB}$ to obtain the state of the `B' subsystem
is depicted as 
\be
	%
  
\ee
Similarly, a non-pure map $\qF$ can be represented as a partially traced out pure map $\hat f$,
\be
\label{equation:Purification}
	%
\ee
The purified $f$ is unique up to a unitary operation applied to the traced-out wire.
This leads to the following lemma,
\begin{equation}
	%
\label{equation:uniquenesspurification}
\end{equation}

\underline{\bf Measurements}.
In the {\bf POVM} formalism (Positive Operator Valued Measure), a measurement $\cM$ is described by a set of `elements' $\{E_x\}_{x\in\cX}$ acting on $\cH$,
with $\cX$ the possible outcomes of the measurement.
These elements satisfy $\sum_{x\in\cX}E_x\dagg E_x={\bf 1}$.
Given a state $\tilde\qr$, the probability of outcome $x$ is given by
$\pr[\cM\to x]_{\tilde\qr}=\tr  E_x\dagg E_x \tilde\qr$; 
diagramatically this is depicted as
\be
	\pr[\cM\to x]_{\qr} \quad=\quad
	%
\ee
The Hermitian conjugate of {\em discarding} equals the preparation of the \textbf{fully mixed state}, up to a factor~$d$.
\be
	%
	= d\cdot\{\mbox{fully mixed state}\}.
\label{FullyMixed}
\ee

\section{Protocols}
\label{sec:protocols}

\subsection{The Quantum One Time Pad}
\label{sec:QOTP}

We showed the QOTP encryption step in (\ref{QOTPthin}).
The full diagram for encryption and decryption of a qubit state
(and including the most general attack
that Eve could launch) is given below, up to a numerical factor~$4$.
\be
	%
\label{equation:QOTPEve}
\ee
The use of thick wires is necessary because (i) Eve's attack $\qF$ is allowed to be very general;
(ii) Alice may not know what state she is encrypting.
The thin spiders connected to thin wires, at the bottom of the diagram, generate the two random key bits.
One copy goes to Alice, one to Bob. 
Eve does not learn the key; this is visually clear, as the attack $\qF$ does not have the key as input.
The encryption by Alice is the doubled version of (\ref{QOTPthin}), with $u$ and $v$ replaced by the random
bits from the thin spiders.
Bob's decryption is the Hermitian conjugate of the encryption. 
{\bf Correctness} of the scheme means that Bob correctly receives the state that Alice sent if
Eve does not interfere. Diagramatically, correctness is proven by removing the $\qF$ box and 
then using spider contraction rules to show that the protocol diagram reduces to a thick wire from Alice to Bob.
\be
	\begin{tikzpicture}
		\begin{pgfonlayer}{nodelayer}
			\node [style=GraySpider] (0) at (-1.25, -4) {};
			\node [style=WHITESPIDER] (1) at (-2, -3.25) {};
			\node [style=none] (2) at (-2, -4) {};
			\node [style=none] (3) at (-2, -2.25) {};
			\node [style=GraySpider] (5) at (-0.5, -4.5) {};
			\node [style=none] (6) at (-2, -3.25) {};
			\node [style=none] (7) at (-0.5, 0) {};
			\node [style=GRAYSPIDER] (8) at (-2, -2.25) {};
			\node [style=WhiteSpider] (9) at (-1.25, -3) {};
			\node [style=WhiteSpider] (10) at (-0.5, -3.5) {};
			\node [style=GraySpider] (11) at (3.5, 2.25) {};
			\node [style=WHITESPIDER] (12) at (1, 3.5) {};
			\node [style=none] (13) at (1, 2.5) {};
			\node [style=none] (14) at (-0.5, 0) {};
			\node [style=none] (18) at (-2, -2.25) {};
			\node [style=GRAYSPIDER] (19) at (1, 2.5) {};
			\node [style=WhiteSpider] (20) at (2.5, 2) {};
			\node [style=WhiteSpider] (21) at (-0.5, -3.5) {};
			\node [style=none] (33) at (-2, -5) {};
			\node [style=none] (34) at (1, 4.25) {};
			\node [style=none] (35) at (0, 4.25) {};
			\node [style=none] (36) at (0, -5.75) {};
			\node [style=none] (37) at (-1, -5.5) {\small Alice};
			\node [style=none] (38) at (0.85, -5.5) {\small Bob};
		\end{pgfonlayer}
		\begin{pgfonlayer}{edgelayer}
			\draw [style=WIRE, bend left=45] (1) to (0);
			\draw [style=WIRE] (3.center) to (2.center);
			\draw (0) to (5);
			\draw [style=WIRE, bend left=45] (8) to (9);
			\draw (9) to (10);
			\draw [style=WIRE, bend left] (12) to (11);
			\draw [style=WIRE, bend right=15] (14.center) to (13.center);
			\draw [style=WIRE, bend left=45] (19) to (20);
			\draw [bend left=15] (20) to (21);
			\draw [bend left=15] (11) to (5);
			\draw [style=WIRE, bend left=15] (18.center) to (14.center);
			\draw [style=WIRE] (19) to (12);
			\draw [style=WIRE] (34.center) to (12);
			\draw [style=new edge style 0] (35.center) to (36.center);
		\end{pgfonlayer}
	\end{tikzpicture}
	\equaltext{(\ref{equation:spiderfusion})}  \begin{tikzpicture}
		\begin{pgfonlayer}{nodelayer}
			\node [style=GraySpider] (0) at (2.75, 0.25) {};
			\node [style=WHITESPIDER] (1) at (0, -1.75) {};
			\node [style=none] (2) at (0, -2.5) {};
			\node [style=none] (3) at (0, -0.75) {};
			\node [style=none] (6) at (0, -1.75) {};
			\node [style=none] (7) at (0, 0.25) {};
			\node [style=GRAYSPIDER] (8) at (0, -0.75) {};
			\node [style=WhiteSpider] (9) at (1.25, 0.5) {};
			\node [style=GraySpider] (11) at (2.75, 0.25) {};
			\node [style=WHITESPIDER] (12) at (0, 2.5) {};
			\node [style=none] (13) at (0, 3.25) {};
			\node [style=none] (14) at (0, 0.25) {};
			\node [style=none] (18) at (0, -0.75) {};
			\node [style=GRAYSPIDER] (19) at (0, 1.5) {};
			\node [style=WhiteSpider] (20) at (1.25, 0.5) {};
			\node [style=none] (21) at (0, -3.5) {};
		\end{pgfonlayer}
		\begin{pgfonlayer}{edgelayer}
			\draw [style=WIRE, bend right, looseness=0.75] (1) to (0);
			\draw [style=WIRE] (3.center) to (2.center);
			\draw [style=WIRE] (7.center) to (6.center);
			\draw [style=WIRE, bend right=15, looseness=0.75] (8) to (9);
			\draw [style=WIRE, bend left] (12) to (11);
			\draw [style=WIRE] (14.center) to (13.center);
			\draw [style=WIRE, bend left=45] (19) to (20);
		\end{pgfonlayer}
	\end{tikzpicture}
	\equaltext{(\ref{equation:spiderfusion})} \begin{tikzpicture}
		\begin{pgfonlayer}{nodelayer}
			\node [style=GraySpider] (0) at (2.75, 0.25) {};
			\node [style=WHITESPIDER] (1) at (0, -0.75) {};
			\node [style=none] (2) at (0, -2.5) {};
			\node [style=none] (3) at (0, -0.75) {};
			\node [style=none] (6) at (0, -1.25) {};
			\node [style=none] (7) at (0, 0.25) {};
			\node [style=GRAYSPIDER] (8) at (0, 0.25) {};
			\node [style=WhiteSpider] (9) at (1.25, 0.5) {};
			\node [style=GraySpider] (11) at (2.75, 0.25) {};
			\node [style=WHITESPIDER] (12) at (0, 1.5) {};
			\node [style=none] (13) at (0, 3.25) {};
			\node [style=none] (14) at (0, 0.25) {};
			\node [style=GRAYSPIDER] (19) at (0, 0.25) {};
			\node [style=WhiteSpider] (20) at (1.25, 0.5) {};
			\node [style=none] (21) at (0, -3.5) {};
		\end{pgfonlayer}
		\begin{pgfonlayer}{edgelayer}
			\draw [style=WIRE, bend right, looseness=0.75] (1) to (0);
			\draw [style=WIRE] (3.center) to (2.center);
			\draw [style=WIRE] (7.center) to (6.center);
			\draw [style=WIRE, bend right=15, looseness=0.75] (8) to (9);
			\draw [style=WIRE, bend left] (12) to (11);
			\draw [style=WIRE] (14.center) to (13.center);
			\draw [style=WIRE, bend left=45] (19) to (20);
		\end{pgfonlayer}
	\end{tikzpicture} 
	\equaltext{(*)}  \frac12 \begin{tikzpicture}
		\begin{pgfonlayer}{nodelayer}
			\node [style=GraySpider] (0) at (1, 0.25) {};
			\node [style=WHITESPIDER] (1) at (0, -0.75) {};
			\node [style=none] (2) at (0, -2.5) {};
			\node [style=none] (3) at (0, -0.75) {};
			\node [style=none] (6) at (0, -1.25) {};
			\node [style=none] (7) at (0, 0.25) {};
			\node [style=GRAYSPIDER] (8) at (0, 0.25) {};
			\node [style=WHITESPIDER] (12) at (0, 1.5) {};
			\node [style=none] (13) at (0, 3.25) {};
			\node [style=none] (14) at (0, 0.25) {};
			\node [style=GRAYSPIDER] (19) at (0, 0.25) {};
			\node [style=none] (21) at (0, -3.5) {};
		\end{pgfonlayer}
		\begin{pgfonlayer}{edgelayer}
			\draw [style=WIRE, bend right, looseness=0.75] (1) to (0);
			\draw [style=WIRE] (3.center) to (2.center);
			\draw [style=WIRE] (7.center) to (6.center);
			\draw [style=WIRE, bend left] (12) to (0);
			\draw [style=WIRE] (14.center) to (13.center);
		\end{pgfonlayer}
	\end{tikzpicture}
	\equaltext{(\ref{equation:boringspider},\ref{equation:spiderfusion})} \frac12 \begin{tikzpicture}
		\begin{pgfonlayer}{nodelayer}
			\node [style=GraySpider] (0) at (1, 0.25) {};
			\node [style=WHITESPIDER] (1) at (0, 0.25) {};
			\node [style=none] (2) at (0, -2.5) {};
			\node [style=none] (3) at (0, -0.25) {};
			\node [style=none] (6) at (0, -1.25) {};
			\node [style=none] (7) at (0, 0.25) {};
			\node [style=WHITESPIDER] (12) at (0, 0.25) {};
			\node [style=none] (13) at (0, 3.25) {};
			\node [style=none] (21) at (0, -3.5) {};
		\end{pgfonlayer}
		\begin{pgfonlayer}{edgelayer}
			\draw [style=WIRE, bend right, looseness=0.75] (1) to (0);
			\draw [style=WIRE] (3.center) to (2.center);
			\draw [style=WIRE] (7.center) to (6.center);
			\draw [style=WIRE, bend left] (12) to (0);
			\draw [style=WIRE] (13.center) to (7.center);
		\end{pgfonlayer}
	\end{tikzpicture}
	\equaltext{(*)} \frac14 \begin{tikzpicture}
		\begin{pgfonlayer}{nodelayer}
			\node [style=none] (2) at (-2, -1.75) {};
			\node [style=none] (23) at (-2, -2.75) {};
			\node [style=none] (24) at (2.25, 2.25) {};
			\node [style=none] (25) at (0, 0) {};
			\node [style=none] (26) at (0, 2.25) {};
			\node [style=none] (27) at (0, -2.75) {};
			\node [style=none] (28) at (-0.9, -2.5) {\small Alice};
			\node [style=none] (29) at (0.75, -2.5) {\small Bob};
		\end{pgfonlayer}
		\begin{pgfonlayer}{edgelayer}
			\draw [style=WIRE, bend left=45] (2.center) to (25.center);
			\draw [style=WIRE, bend right=45] (25.center) to (24.center);
			\draw [style=new edge style 0] (26.center) to (27.center);
		\end{pgfonlayer}
	\end{tikzpicture}
\label{QOTPcorrect}
\ee
Step (*) consists of two parts:
First, a double thin connection between complementary thin spiders yields a numerical factor $1/2$ and a disconnect,\footnote{
We ignore a subtlety involving so-called `antipodes' \cite{Coecke2017}, which is allowed here because the antipode evaluates to
the identity in the case of the combination of the $x$ and $z$ basis.
}
as we saw in (\ref{equation:ComplementarityPositive}); now we have a version of this situation where one of the spiders is doubled,
yielding a factor $1/4$ instead of $1/2$.
Second, the isolated thin spider yields a factor $2$ due to the basis summation.

{\bf Security} of the scheme means that Eve does not learn what Alice's state is.
Diagrammatically, security is `proven' by tracing out Bob and showing that the output of Eve's 
attack $\qF$ is decoupled from Alice's input wire.
Tracing out Bob removes the entire branch going from Alice to Bob.
The part of the diagram that serves as input to $\qF$ reduces to
\begin{equation}
		\begin{tikzpicture}
			\begin{pgfonlayer}{nodelayer}
				\node [style=GraySpider] (0) at (1, -0.75) {};
				\node [style=WHITESPIDER] (1) at (0, 0.25) {};
				\node [style=none] (2) at (0, -1.75) {};
				\node [style=none] (3) at (0, 1.25) {};
				\node [style=none] (6) at (0, 0.25) {};
				\node [style=none] (7) at (0, 2.25) {};
				\node [style=GRAYSPIDER] (8) at (0, 1.25) {};
				\node [style=WhiteSpider] (9) at (1.75, 0.25) {};
				\node [style=none] (11) at (0, -2.5) {};
			\end{pgfonlayer}
			\begin{pgfonlayer}{edgelayer}
				\draw [style=WIRE, bend left=45] (1) to (0);
				\draw [style=WIRE] (3.center) to (2.center);
				\draw [style=WIRE] (7.center) to (6.center);
				\draw [style=WIRE, out=0, in=100] (8) to (9);
			\end{pgfonlayer}
		\end{tikzpicture} 
		\equaltext{(\ref{FullyMixed})} \begin{tikzpicture}
			\begin{pgfonlayer}{nodelayer}
				\node [style=WHITESPIDER] (1) at (0, 0.25) {};
				\node [style=none] (2) at (0, -1.75) {};
				\node [style=none] (3) at (0, 1.25) {};
				\node [style=none] (6) at (0, 0.25) {};
				\node [style=none] (7) at (0, 2.25) {};
				\node [style=GRAYSPIDER] (8) at (0, 1.25) {};
				
				\node [style=none] (12) at (3.25, 0.25) {};
				\node [style=none] (13) at (2, 0.25) {};
				\node [style=none] (14) at (3, 0) {};
				\node [style=none] (15) at (2.25, 0) {};
				\node [style=none] (16) at (2.75, -0.25) {};
				\node [style=none] (17) at (2.5, -0.25) {};
				
				\node [style=none] (18) at (1.625, -0.75) {};
				\node [style=none] (19) at (0.375, -0.75) {};
				\node [style=none] (20) at (1.375, -1) {};
				\node [style=none] (21) at (0.625, -1) {};
				\node [style=none] (22) at (1.125, -1.25) {};
				\node [style=none] (23) at (0.875, -1.25) {};
				
				\node [style=none] (24) at (1, -0.75) {};
				\node [style=none] (25) at (2.625, 0.25) {};
				\node [style=none] (11) at (0, -2.5) {};
			\end{pgfonlayer}
			\begin{pgfonlayer}{edgelayer}
				\draw [style=WIRE] (3.center) to (2.center);
				\draw [style=WIRE] (7.center) to (6.center);
				\draw [style=WIRE] (12.center) to (13.center);
				\draw [style=WIRE] (14.center) to (15.center);
				\draw [style=WIRE] (16.center) to (17.center);
				\draw [style=WIRE] (18.center) to (19.center);
				\draw [style=WIRE] (20.center) to (21.center);
				\draw [style=WIRE] (22.center) to (23.center);
				\draw [style=WIRE, bend right=330] (8) to (25.center);
				\draw [style=WIRE, bend right=315, looseness=1.25] (6.center) to (24.center);
			\end{pgfonlayer}
		\end{tikzpicture} 
		\equaltext{(\ref{FullyMixed})} \begin{tikzpicture}
			\begin{pgfonlayer}{nodelayer}
				\node [style=WHITESPIDER] (1) at (0, 0.25) {};
				\node [style=none] (2) at (0, -1.75) {};
				\node [style=none] (3) at (0, 1.25) {};
				\node [style=none] (6) at (0, 0.25) {};
				\node [style=none] (7) at (0, 2.25) {};
				\node [style=GRAYSPIDER] (8) at (0, 1.25) {};
				\node [style=none] (24) at (1, -0.75) {};
				\node [style=none] (25) at (1.975, 0.25) {};
				\node [style=WhiteSpider] (26) at (1, -0.75) {};
				\node [style=GraySpider] (27) at (2, 0.25) {};
				\node [style=none] (11) at (0, -2.5) {};
			\end{pgfonlayer}
			\begin{pgfonlayer}{edgelayer}
				\draw [style=WIRE] (3.center) to (2.center);
				\draw [style=WIRE] (7.center) to (6.center);
				\draw [style=WIRE, bend right=330] (8) to (25.center);
				\draw [style=WIRE, bend right=315, looseness=1.25] (6.center) to (24.center);
			\end{pgfonlayer}
		\end{tikzpicture} 
		\equaltext{(\ref{equation:bastardspiderfusion})} \begin{tikzpicture}
			\begin{pgfonlayer}{nodelayer}
				\node [style=none] (2) at (0, -1.75) {};
				\node [style=none] (3) at (0, 1.25) {};
				\node [style=none] (7) at (0, 2.25) {};
				\node [style=GraySpider] (12) at (0, 1.25) {};
				\node [style=WhiteSpider] (13) at (0, 0.25) {};
			\end{pgfonlayer}
			\begin{pgfonlayer}{edgelayer}
				\draw [style=WIRE] (3.center) to (2.center);
				\draw [style=WIRE] (7.center) to (3.center);
			\end{pgfonlayer}
		\end{tikzpicture} 
		\equaltext{(\ref{equation:ComplementarityPositive})} \frac{1}{2} \begin{tikzpicture}
			\begin{pgfonlayer}{nodelayer}
				\node [style=none] (2) at (0, -1.75) {};
				\node [style=none] (3) at (0, 1.25) {};
				\node [style=none] (7) at (0, 2.25) {};
				\node [style=GraySpider] (12) at (0, 1.25) {};
				\node [style=WhiteSpider] (13) at (0, 0.25) {};
				\node [style=none] (14) at (0, -2.5) {};
			\end{pgfonlayer}
			\begin{pgfonlayer}{edgelayer}
				\draw [style=WIRE] (7.center) to (3.center);
				\draw [style=WIRE] (13) to (2.center);
			\end{pgfonlayer}
		\end{tikzpicture} 
		\equaltext{(\ref{FullyMixed})} \frac{1}{2} \begin{tikzpicture}
			\begin{pgfonlayer}{nodelayer}
				\node [style=none] (2) at (0, -1.75) {};
				\node [style=none] (3) at (0, 1.425) {};
				\node [style=none] (7) at (0, 2.25) {};
				\node [style=none] (14) at (0, -2.5) {};
				\node [style=none] (15) at (-0.625, 1.425) {};
				\node [style=none] (16) at (0.625, 1.425) {};
				\node [style=none] (17) at (-0.375, 1.175) {};
				\node [style=none] (18) at (0.375, 1.175) {};
				\node [style=none] (19) at (-0.125, 0.925) {};
				\node [style=none] (20) at (0.125, 0.925) {};
				\node [style=none] (21) at (0, -0.075) {};
				\node [style=none] (22) at (0.625, -0.075) {};
				\node [style=none] (23) at (-0.625, -0.075) {};
				\node [style=none] (24) at (0.375, 0.175) {};
				\node [style=none] (25) at (-0.375, 0.175) {};
				\node [style=none] (26) at (0.125, 0.425) {};
				\node [style=none] (27) at (-0.125, 0.425) {};
			\end{pgfonlayer}
			\begin{pgfonlayer}{edgelayer}
				\draw [style=WIRE] (7.center) to (3.center);
				\draw [style=WIRE] (15.center) to (16.center);
				\draw [style=WIRE] (17.center) to (18.center);
				\draw [style=WIRE] (19.center) to (20.center);
				\draw [style=WIRE] (22.center) to (23.center);
				\draw [style=WIRE] (24.center) to (25.center);
				\draw [style=WIRE] (26.center) to (27.center);
				\draw [style=WIRE] (21.center) to (2.center);
			\end{pgfonlayer}
		\end{tikzpicture}
\label{disconnect8}
\end{equation}
which indeed entirely disconnects Eve from Alice's state.

\subsection{Teleportation}
\label{sec:teleport}

Alice holds a qubit state $\ket\qj$.
Alice and Bob share an EPR qubit pair.
Alice performs a measurement on her qubit $\ket\qj$ and her half of the EPR pair,
in the Bell basis; this results in a two-bit measurement outcome $(u,v)$.
The two bits are communicated to Bob.
Bob applies the Bell operation $(u,v)$, which is essentially the decryption $\qs_x^v \qs_z^u$,
to his half of the EPR pair. The end result is that Bob's qubit has state $\ket\qj$.

Measurement in the Bell basis is the Hermitian conjugate of the Bell state preparation (\ref{BellStateuv})
and can be depicted as
\be
	\begin{tikzpicture}
		\begin{pgfonlayer}{nodelayer}
			\node [style=none] (0) at (-1, -1.5) {};
			\node [style=none] (3) at (-1, -0.5) {};
			\node [style=none] (9) at (-2, -1.5) {};
			\node [style=GraySpider] (15) at (-2, 0.5) {};
			\node [style=WhiteSpider] (18) at (-1, 0.5) {};
			\node [style=GRAYSPIDER] (19) at (-1, -0.5) {};
			\node [style=WHITESPIDER] (20) at (-2, -0.5) {};
			\node [style=none] (30) at (-2,1.5) {};
			\node [style=none] (31) at (-1,1.5) {};
		\end{pgfonlayer}
		\begin{pgfonlayer}{edgelayer}
			\draw [style=WIRE] (3.center) to (0.center);
			\draw [style=WIRE] (19) to (20);
			\draw [style=WIRE] (18) to (19);
			\draw [style=WIRE] (15) to (20);
			\draw [style=WIRE] (9.center) to (20);
			\draw (15) to (30);
			\draw (18) to (31);
		\end{pgfonlayer}
	\end{tikzpicture}
\label{BellMeas}
\ee
The thin wires carry the bits $(u,v)$. 
The full diagram for teleportation of a state $\qr$ is
\begin{equation}
	\begin{tikzpicture}
		\begin{pgfonlayer}{nodelayer}
			\node [style=none] (0) at (-1, -1.5) {};
			\node [style=none] (1) at (1, -1.5) {};
			\node [style=none] (2) at (1, 1.75) {};
			\node [style=none] (3) at (-1, -0.5) {};
			\node [style=WHITESPIDER] (4) at (-1, -0.5) {};
			\node [style=GRAYSPIDER] (5) at (-2, -0.5) {};
			\node [style=GraySpider] (6) at (-1, 0.5) {};
			\node [style=WhiteSpider] (7) at (-2, 0.5) {};
			\node [style=none] (9) at (-2, -2.25) {};
			\node [style=WHITESPIDER] (10) at (1, 1.75) {};
			\node [style=GRAYSPIDER] (11) at (1, 2.75) {};
			\node [style=GraySpider] (12) at (0, 1.75) {};
			\node [style=WhiteSpider] (13) at (0, 2.75) {};
			\node [style=none] (14) at (1, 3.75) {};
			\node [style=GraySpider] (15) at (-2, 0.5) {};
			\node [style=GraySpider] (16) at (0, 2.75) {};
			\node [style=WhiteSpider] (17) at (0, 1.75) {};
			\node [style=WhiteSpider] (18) at (-1, 0.5) {};
			\node [style=GRAYSPIDER] (19) at (-1, -0.5) {};
			\node [style=WHITESPIDER] (20) at (-2, -0.5) {};
			\node [style=WHITESPIDER] (21) at (1, 2.75) {};
			\node [style=GRAYSPIDER] (22) at (1, 1.75) {};
			\node [style=none] (23) at (-0.5, 4.5) {};
			\node [style=none] (24) at (-0.5, -2.3) {};
			\node [style=none] (25) at (0.25, 4.25) {Bob};
			\node [style=none] (26) at (-1.4, 4.25) {Alice};
			\node [style=none] (27) at (-2, -3) {};
			\node [style=none] (28) at (-2.4, -2) {\small $\rho$};
			\node [style=none] (28) at (1.3,3.6) {\small $\rho$};
		\end{pgfonlayer}
		\begin{pgfonlayer}{edgelayer}
			\draw [style=WIRE, bend right=90, looseness=1.25] (0.center) to (1.center);
			\draw [style=WIRE] (2.center) to (1.center);
			\draw [style=WIRE] (3.center) to (0.center);
			\draw [style=WIRE] (4) to (5);
			\draw [style=WIRE] (6) to (4);
			\draw [style=WIRE] (7) to (5);
			\draw [style=WIRE] (9.center) to (5);
			\draw [style=WIRE] (11) to (10);
			\draw [style=WIRE] (10) to (12);
			\draw [style=WIRE] (11) to (13);
			\draw [bend right=45] (12) to (6);
			\draw [bend right=45] (13) to (7);
			\draw [style=WIRE] (14.center) to (11);
			\draw [style=new edge style 0] (23.center) to (24.center);
		\end{pgfonlayer}
	\end{tikzpicture}
\end{equation}
(Here the thick-wired `cup' represents the creation of the EPR state $\ket{\qF_{00}}$, 
i.e.~we have omitted the spider that creates the EPR state.)
By yanking the path of $\qr$ (thick wire) to a straight line, the above diagram can be deformed such
that it becomes precisely the leftmost picture in (\ref{QOTPcorrect}).
Thus working with diagrams makes the equivalence between Quantum On-Time Padding and teleport 
particularly easy to visualise.

\subsection{Quantum Key Distribution}
\label{sec:QKD}

We briefly discuss Kissinger, Tull and Westerbaan's diagrammatic treatment \cite{KTW2017} of BB84  \cite{BB84}.
The main component of BB84 is the use of the quantum channel:
Alice encodes a random bit $x$ in a randomly chosen basis (standard basis or complementary)
and sends it to Bob.
Bob also selects one of these two bases at random and measures in that basis.
Only the events where Bob's basis matches Alice's are relevant.
Before reaching Bob, the qubit may be manipulated by Eve.
Eve's attack is allowed to be anything at all, i.e. Eve may entangle her own quantum system with the qubit,
and use quantum memory and quantum computation.
The diagram for a single channel use (in the standard basis) is as follows.
\be
	\begin{tikzpicture}
		\begin{pgfonlayer}{nodelayer}
			\node [style=ASYM BOX] (0) at (0, 0) {\reflectbox{$\qF$}};
			\node [style=none] (1) at (-0.75, 0) {};
			\node [style=none] (2) at (0, 0) {};
			\node [style=none] (3) at (-0.75, 1.25) {};
			\node [style=none] (4) at (-0.5, 1.25) {};
			\node [style=none] (5) at (0, -1.25) {};
			\node [style=WhiteSpider] (9) at (0.5, 1.25) {};
			\node [style=WhiteSpider] (10) at (0, -1.25) {};
			\node [style=none] (11) at (0.5, 2) {};
			\node [style=none] (12) at (0, -2) {};
			\node [style=none] (13) at (-0.25, 1.25) {};
			\node [style=none] (14) at (-1.25, 1.25) {};
			\node [style=none] (15) at (-0.5, 1.5) {};
			\node [style=none] (16) at (-1, 1.5) {};
			\node [style=none] (17) at (-0.65, 1.75) {};
			\node [style=none] (18) at (-0.65, 1.75) {};
			\node [style=none] (19) at (-0.85, 1.75) {};
			\node [style=none] (20) at (0.5, 0) {};
			\node [style=none] (30) at (0.8,-1.8) {\small Alice};
			\node [style=none] (31) at (1.2, 1.8) {\small Bob};
			\node [style=none] (32) at (-1, 1.8) {\small Eve};
		\end{pgfonlayer}
		\begin{pgfonlayer}{edgelayer}
			\draw [style=WIRE] (3.center) to (1.center);
			\draw [style=WIRE] (2.center) to (5.center);
			\draw (11.center) to (9);
			\draw (10) to (12.center);
			\draw [style=WIRE] (9) to (20.center);
		\end{pgfonlayer}
	\end{tikzpicture}
\ee
Here $\qF$ stands for the attack.
The same diagram, but with grey spiders, applies to the other basis.
{\em After} the attack $\qF$, Eve learns the basis choice and tries to exploit her quantum side information
to guess the data bit sent by Alice.
The statement that the protocol is secure\footnote{
Here we show the noiseless case only. Noise is handled as in \cite{KTW2017}.
}
is formulated diagramatically as follows (Theorem~3.1 in \cite{KTW2017}):
\be
	\begin{tikzpicture}
		\begin{pgfonlayer}{nodelayer}
			\node [style=ASYM BOX] (0) at (0, 0) {$\Phi$};
			\node [style=none] (1) at (-0.75, 0) {};
			\node [style=none] (2) at (0, 0) {};
			\node [style=none] (3) at (-0.75, 1.25) {};
			\node [style=none] (4) at (-0.5, 1.25) {};
			\node [style=none] (5) at (0, -1.25) {};
			\node [style=WhiteSpider] (9) at (0.5, 1.25) {};
			\node [style=WhiteSpider] (10) at (0, -1.25) {};
			\node [style=none] (11) at (0.5, 2) {};
			\node [style=none] (12) at (0, -2) {};
			\node [style=none] (13) at (-0.25, 1.25) {};
			\node [style=none] (14) at (-1.25, 1.25) {};
			\node [style=none] (15) at (-0.5, 1.5) {};
			\node [style=none] (16) at (-1, 1.5) {};
			\node [style=none] (17) at (-0.65, 1.75) {};
			\node [style=none] (18) at (-0.65, 1.75) {};
			\node [style=none] (19) at (-0.85, 1.75) {};
			\node [style=none] (20) at (0.5, 0) {};
		\end{pgfonlayer}
		\begin{pgfonlayer}{edgelayer}
			\draw [style=WIRE] (3.center) to (1.center);
			\draw [style=WIRE] (2.center) to (5.center);
			\draw (11.center) to (9);
			\draw (10) to (12.center);
			\draw [style=WIRE] (18.center) to (19.center);
			\draw [style=WIRE] (15.center) to (16.center);
			\draw [style=WIRE] (13.center) to (14.center);
			\draw [style=WIRE] (9) to (20.center);
		\end{pgfonlayer}
	\end{tikzpicture} = \begin{tikzpicture}
	\begin{pgfonlayer}{nodelayer}
		\node [style=none] (0) at (0, 1) {};
		\node [style=none] (1) at (0, -1) {};
	\end{pgfonlayer}
	\begin{pgfonlayer}{edgelayer}
		\draw (0.center) to (1.center);
	\end{pgfonlayer}
\end{tikzpicture} ~~~~~ \land ~~~~~ \begin{tikzpicture}
	\begin{pgfonlayer}{nodelayer}
		\node [style=ASYM BOX] (0) at (0, 0) {$\Phi$};
		\node [style=none] (1) at (-0.75, 0) {};
		\node [style=none] (2) at (0, 0) {};
		\node [style=none] (3) at (-0.75, 1.25) {};
		\node [style=none] (4) at (-1.25, 1.25) {};
		\node [style=none] (5) at (0, -1.25) {};
		\node [style=WhiteSpider] (9) at (0.5, 1.25) {};
		\node [style=WhiteSpider] (10) at (0, -1.25) {};
		\node [style=none] (11) at (0.5, 2) {};
		\node [style=none] (12) at (0, -2) {};
		\node [style=none] (13) at (-0.25, 1.25) {};
		\node [style=none] (14) at (-1.25, 1.25) {};
		\node [style=none] (15) at (-0.5, 1.5) {};
		\node [style=none] (16) at (-1, 1.5) {};
		\node [style=none] (17) at (-0.65, 1.75) {};
		\node [style=none] (18) at (-0.65, 1.75) {};
		\node [style=none] (19) at (-0.85, 1.75) {};
		\node [style=GraySpider] (20) at (0.5, 1.25) {};
		\node [style=GraySpider] (21) at (0, -1.25) {};
		\node [style=none] (22) at (0.5, 0) {};
	\end{pgfonlayer}
	\begin{pgfonlayer}{edgelayer}
		\draw [style=WIRE] (3.center) to (1.center);
		\draw [style=WIRE] (2.center) to (5.center);
		\draw (11.center) to (9);
		\draw (10) to (12.center);
		\draw [style=WIRE] (18.center) to (19.center);
		\draw [style=WIRE] (15.center) to (16.center);
		\draw [style=WIRE] (13.center) to (14.center);
		\draw [style=WIRE] (20) to (22.center);
	\end{pgfonlayer}
\end{tikzpicture} = \begin{tikzpicture}
\begin{pgfonlayer}{nodelayer}
	\node [style=none] (0) at (0, 1) {};
	\node [style=none] (1) at (0, -1) {};
\end{pgfonlayer}
\begin{pgfonlayer}{edgelayer}
	\draw (0.center) to (1.center);
\end{pgfonlayer}
\end{tikzpicture}
	\quad\quad\implies\quad\quad
	\exists_\qr \begin{tikzpicture}
		\begin{pgfonlayer}{nodelayer}
			\node [style=ASYM BOX] (0) at (0, 0) {$\Phi$};
			\node [style=none] (1) at (0.5, 0) {};
			\node [style=none] (2) at (0, 0) {};
			\node [style=none] (3) at (0.5, 1.25) {};
			\node [style=none] (4) at (-0.5, 1.25) {};
			\node [style=none] (5) at (0, -1.25) {};
			\node [style=none] (6) at (-1, 1) {\tiny {\rm Eve}};
			\node [style=none] (7) at (1, 1) {\tiny {\rm Bob}};
			\node [style=none] (8) at (-0.75, -1) {\tiny {\rm Alice}};
			\node [style=none] (9) at (-0.5, 0) {};
		\end{pgfonlayer}
		\begin{pgfonlayer}{edgelayer}
			\draw [style=WIRE] (3.center) to (1.center);
			\draw [style=WIRE] (2.center) to (5.center);
			\draw [style=WIRE] (4.center) to (9.center);
		\end{pgfonlayer}
	\end{tikzpicture} = \begin{tikzpicture}
	\begin{pgfonlayer}{nodelayer}
		\node [style=STATERIGHT] (0) at (0, 0) {\rotatebox{90}{$\rho$}};
		\node [style=none] (1) at (0, 1) {};
		\node [style=none] (2) at (0, -0.5) {};
		\node [style=none] (3) at (-0.1, 1.2) {\tiny {\rm Eve}};
	\end{pgfonlayer}
	\begin{pgfonlayer}{edgelayer}
		\draw [style=WIRE] (1.center) to (2.center);
	\end{pgfonlayer}
\end{tikzpicture}  ~~~ \begin{tikzpicture}
\begin{pgfonlayer}{nodelayer}
	\node [style=none] (0) at (0, 1) {};
	\node [style=none] (1) at (0, -1.25) {};
	\node [style=none] (2) at (0.6, -1.1) {\tiny {\rm Alice}};
	\node [style=none] (3) at (0.5, 0.9) {\tiny {\rm Bob}};
\end{pgfonlayer}
\begin{pgfonlayer}{edgelayer}
	\draw [style=WIRE] (0.center) to (1.center);
\end{pgfonlayer}
\end{tikzpicture}
\label{QKDsecurity}
\ee
The two diagrams to the left of the implication arrow say that Alice and Bob see no disturbance in the two bases
that they are using. The right hand side says that Eve is decoupled from the information that is going from Alice to Bob.

The proof of (\ref{QKDsecurity}) consists of two parts. 
First, from the left hand side (non-disturbance) a rule is derived that allows one to pull
spiders through the purification of $\qF$; here it is important that the rule applies only to those bases
that are observed to have no disturbance, i.e.\;in BB84 the $z$-basis and $x$-basis.
Second, an additional disconnected branch is pasted to Bob's wire and then all the spiders involved with Bob's wire are
pulled through~$\qF$.
Here we will not write out the full proof from \cite{KTW2017}; we highlight only one part of it 
that we will re-use in the diagrammatic reduction in Section~\ref{sec:8state}. 
\begin{lemma}
\label{lemma:existsUpsi}
Let the white spider represent any orthonormal basis. 
Let $V$ be a purification of $\qF$.
It holds that
\be
	\begin{tikzpicture}
		\begin{pgfonlayer}{nodelayer}
			\node [style=ASYM BOX] (0) at (0, 0) {$\Phi$};
			\node [style=none] (1) at (-0.75, 0) {};
			\node [style=none] (2) at (0, 0) {};
			\node [style=none] (3) at (-0.75, 1.25) {};
			\node [style=none] (4) at (-0.5, 1.25) {};
			\node [style=none] (5) at (0, -1.25) {};
			\node [style=WhiteSpider] (9) at (0.5, 1.25) {};
			\node [style=WhiteSpider] (10) at (0, -1.25) {};
			\node [style=none] (11) at (0.5, 2) {};
			\node [style=none] (12) at (0, -2) {};
			\node [style=none] (13) at (-0.25, 1.25) {};
			\node [style=none] (14) at (-1.25, 1.25) {};
			\node [style=none] (15) at (-0.5, 1.5) {};
			\node [style=none] (16) at (-1, 1.5) {};
			\node [style=none] (17) at (-0.65, 1.75) {};
			\node [style=none] (18) at (-0.65, 1.75) {};
			\node [style=none] (19) at (-0.85, 1.75) {};
			\node [style=none] (20) at (0.5, 0) {};
		\end{pgfonlayer}
		\begin{pgfonlayer}{edgelayer}
			\draw [style=WIRE] (3.center) to (1.center);
			\draw [style=WIRE] (2.center) to (5.center);
			\draw (11.center) to (9);
			\draw (10) to (12.center);
			\draw [style=WIRE] (18.center) to (19.center);
			\draw [style=WIRE] (15.center) to (16.center);
			\draw [style=WIRE] (13.center) to (14.center);
			\draw [style=WIRE] (9) to (20.center);
		\end{pgfonlayer}
	\end{tikzpicture}= \begin{tikzpicture}
	\begin{pgfonlayer}{nodelayer}
		\node [style=none] (0) at (0, 1) {};
		\node [style=none] (1) at (0, -1) {};
	\end{pgfonlayer}
	\begin{pgfonlayer}{edgelayer}
		\draw (0.center) to (1.center);
	\end{pgfonlayer}
\end{tikzpicture} 
	\quad\quad\implies\quad\quad
	\exists_{\qj, {\rm unitary}\,U} \quad \begin{tikzpicture}
		\begin{pgfonlayer}{nodelayer}
			\node [style=asym box] (0) at (0, 0) {\reflectbox{V}};
			\node [style=none] (1) at (0, -1.5) {};
			\node [style=none] (11) at (-0.5, 0) {};
			\node [style=none] (12) at (-0.5, 1.5) {};
			\node [style=none] (13) at (0.35, 0) {};
			\node [style=none] (14) at (0.35, 1.5) {};
		\end{pgfonlayer}
		\begin{pgfonlayer}{edgelayer}
			\draw (13.center) to (14.center);
			\draw (12.center) to (11.center);
			\draw (0) to (1);
		\end{pgfonlayer}
	\end{tikzpicture}
	= \begin{tikzpicture}
		\begin{pgfonlayer}{nodelayer}
			\node [style=none] (0) at (0.8, 2) {};
			\node [style=none] (1) at (0.8, -2) {};
			\node [style=WhiteSpider] (2) at (0.8, -1) {};
			\node [style=medium box] (3) at (-1.2, 0.25) {U};
			\node [style=StateMid] (4) at (-1.625, -1.5) {\rotatebox{-45}{$\psi$}};
			\node [style=none] (5) at (-1.95, 0.25) {};
			\node [style=none] (6) at (-1.875, 0.75) {};
			\node [style=none] (7) at (-1.2, 0.75) {};
			\node [style=none] (8) at (-0.7, 0.75) {};
			\node [style=none] (9) at (-0.45, 0.25) {};
			\node [style=none] (10) at (-1.375, -2) {};
			\node [style=none] (11) at (-1.925, 0.75) {};
			\node [style=none] (12) at (-1.7, -2) {};
			\node [style=none] (13) at (-0.5, 2) {};
			\node [style=none] (14) at (-1.25, 2) {};
			\node [style=none] (15) at (-2, 2) {};
			\node [style=none] (16) at (-0.5, 0.75) {};
			\node [style=none] (17) at (-1.25, 0.75) {};
			\node [style=none] (18) at (-2, 0.75) {};
			\node [style=WhiteSpider] (19) at (-2, 2) {};
			\node [style=WhiteSpider] (20) at (-0.5, 2) {};
		\end{pgfonlayer}
		\begin{pgfonlayer}{edgelayer}
			\draw (0.center) to (1.center);
			\draw [bend left=45, looseness=1.25] (2) to (8.center);
			\draw [bend left=15] (7.center) to (10.center);
			\draw [bend right=15] (11.center) to (12.center);
			\draw (15.center) to (18.center);
			\draw (14.center) to (17.center);
			\draw (13.center) to (16.center);
		\end{pgfonlayer}
	\end{tikzpicture}
\label{existsUpsi}
\ee
\underline{\em Proof}:
See the proof of Theorem~3.1 in \cite{KTW2017}.
\hfill$\square$

\end{lemma}

\begin{lemma}
\label{lemma:pullthrough}
Let the white spider represent any orthonormal basis. 
Let $V$ be a purification of $\qF$.
It holds that
\be
	\begin{tikzpicture}
		\begin{pgfonlayer}{nodelayer}
			\node [style=ASYM BOX] (0) at (0, 0) {$\Phi$};
			\node [style=none] (1) at (-0.75, 0) {};
			\node [style=none] (2) at (0, 0) {};
			\node [style=none] (3) at (-0.75, 1.25) {};
			\node [style=none] (4) at (-0.5, 1.25) {};
			\node [style=none] (5) at (0, -1.25) {};
			\node [style=WhiteSpider] (9) at (0.5, 1.25) {};
			\node [style=WhiteSpider] (10) at (0, -1.25) {};
			\node [style=none] (11) at (0.5, 2) {};
			\node [style=none] (12) at (0, -2) {};
			\node [style=none] (13) at (-0.25, 1.25) {};
			\node [style=none] (14) at (-1.25, 1.25) {};
			\node [style=none] (15) at (-0.5, 1.5) {};
			\node [style=none] (16) at (-1, 1.5) {};
			\node [style=none] (17) at (-0.65, 1.75) {};
			\node [style=none] (18) at (-0.65, 1.75) {};
			\node [style=none] (19) at (-0.85, 1.75) {};
			\node [style=none] (20) at (0.5, 0) {};
		\end{pgfonlayer}
		\begin{pgfonlayer}{edgelayer}
			\draw [style=WIRE] (3.center) to (1.center);
			\draw [style=WIRE] (2.center) to (5.center);
			\draw (11.center) to (9);
			\draw (10) to (12.center);
			\draw [style=WIRE] (18.center) to (19.center);
			\draw [style=WIRE] (15.center) to (16.center);
			\draw [style=WIRE] (13.center) to (14.center);
			\draw [style=WIRE] (9) to (20.center);
		\end{pgfonlayer}
	\end{tikzpicture}= \begin{tikzpicture}
	\begin{pgfonlayer}{nodelayer}
		\node [style=none] (0) at (0, 1) {};
		\node [style=none] (1) at (0, -1) {};
	\end{pgfonlayer}
	\begin{pgfonlayer}{edgelayer}
		\draw (0.center) to (1.center);
	\end{pgfonlayer}
\end{tikzpicture}
	\quad\quad\implies\quad\quad
	\begin{tikzpicture}
		\begin{pgfonlayer}{nodelayer}
			\node [style=none] (0) at (0.025, 1.25) {};
			\node [style=none] (1) at (-0.975, 2.25) {};
			\node [style=none] (3) at (-0.275, -1.75) {};
			\node [style=WhiteSpider] (5) at (0.025, 1.25) {};
			\node [style=asym box] (9) at (-0.275, -0.25) {\reflectbox{$V$}};
			\node [style=none] (14) at (1.125, 2.25) {};
			\node [style=none] (15) at (-0.725, -0.25) {};
			\node [style=none] (16) at (-1.725, 2.25) {};
			\node [style=none] (17) at (0, -0.25) {};
		\end{pgfonlayer}
		\begin{pgfonlayer}{edgelayer}
			\draw [bend right] (5) to (14.center);
			\draw [bend left] (5) to (1.center);
			\draw [bend left=15] (15.center) to (16.center);
			\draw (17.center) to (5);
			\draw (9) to (3.center);
		\end{pgfonlayer}
	\end{tikzpicture} = \begin{tikzpicture}
	\begin{pgfonlayer}{nodelayer}
		\node [style=none] (1) at (-0.75, 1) {};
		\node [style=none] (3) at (0, -1.25) {};
		\node [style=WhiteSpider] (5) at (0, 0) {};
		\node [style=asym box] (9) at (-0.75, 1.75) {\reflectbox{$V$}};
		\node [style=none] (14) at (1.375, 1.25) {};
		\node [style=none] (15) at (-1.5, 1.75) {};
		\node [style=none] (16) at (-1.5, 2.75) {};
		\node [style=none] (17) at (-0.5, 2.75) {};
		\node [style=none] (18) at (-0.5, 1.75) {};
		\node [style=none] (19) at (1.375, 2.75) {};
	\end{pgfonlayer}
	\begin{pgfonlayer}{edgelayer}
		\draw [bend right=45] (5) to (14.center);
		\draw [bend left] (5) to (1.center);
		\draw (3.center) to (5);
		\draw (15.center) to (16.center);
		\draw (17.center) to (18.center);
		\draw (9) to (1.center);
		\draw (19.center) to (14.center);
	\end{pgfonlayer}
\end{tikzpicture}
	\quad\wedge\quad
	\begin{tikzpicture}
		\begin{pgfonlayer}{nodelayer}
			\node [style=none] (0) at (0.75, 0) {};
			\node [style=none] (1) at (1.75, -1.25) {};
			\node [style=WhiteSpider] (5) at (0.75, 0) {};
			\node [style=asym box] (9) at (-0.5, -1.75) {\reflectbox{V}};
			\node [style=none] (14) at (-0.35, -1.75) {};
			\node [style=none] (15) at (-0.5, -3.25) {};
			\node [style=none] (16) at (-1, -1.75) {};
			\node [style=none] (17) at (-1, 1.25) {};
			\node [style=none] (18) at (0.75, 1.25) {};
			\node [style=none] (19) at (1.75, -3.25) {};
		\end{pgfonlayer}
		\begin{pgfonlayer}{edgelayer}
			\draw [bend right=45] (5) to (14.center);
			\draw [bend left] (5) to (1.center);
			\draw (16.center) to (17.center);
			\draw (18.center) to (5);
			\draw (9) to (15.center);
			\draw (1.center) to (19.center);
		\end{pgfonlayer}
	\end{tikzpicture}
	= \begin{tikzpicture}
		\begin{pgfonlayer}{nodelayer}
			\node [style=none] (0) at (0, -1) {};
			\node [style=none] (1) at (-1, -2) {};
			\node [style=none] (3) at (-0.5, 2) {};
			\node [style=WhiteSpider] (5) at (0, -1) {};
			\node [style=asym box] (9) at (0, 0.5) {\reflectbox{V}};
			\node [style=none] (14) at (1.1, -2) {};
			\node [style=none] (15) at (0.35, 0.5) {};
			\node [style=none] (16) at (0.35, 2) {};
			\node [style=none] (17) at (-0.5, 0.5) {};
		\end{pgfonlayer}
		\begin{pgfonlayer}{edgelayer}
			\draw [bend left] (5) to (14.center);
			\draw [bend right] (5) to (1.center);
			\draw (15.center) to (16.center);
			\draw (3.center) to (17.center);
			\draw (9) to (5);
		\end{pgfonlayer}
	\end{tikzpicture}
\label{pullthrough}
\ee
\end{lemma}
\underline{\em Proof}:
See the proof of Theorem~3.1 in \cite{KTW2017}.
\hfill$\square$

(If $\qF$ is not pure then the left output wire of $V$ represents a space that is larger than Eve's Hilbert space;
a part has to be traced out to obtain Eve's space.)
Note that the third equation in (\ref{pullthrough}) follows from the second equation by yanking down the rightmost wire.

\subsection{Eight-state encoding}
\label{sec:8state}

\begin{figure}[h!]
	\begin{center}
	\includegraphics[width=0.3\linewidth]{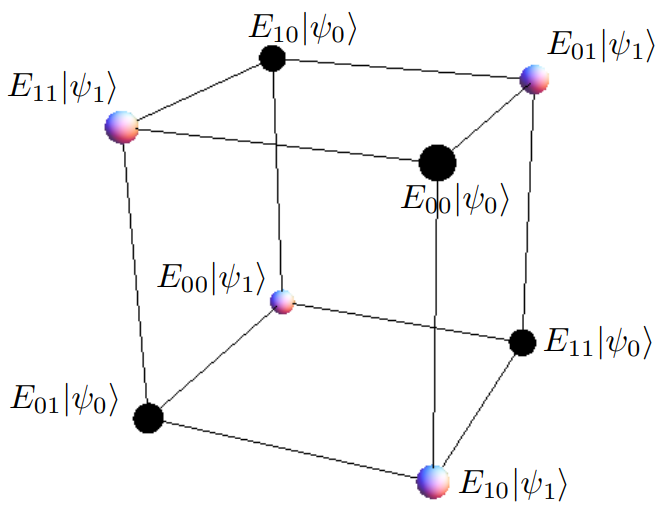}
\caption{\it The eight cipherstates $E_{uv}\ket{\psi_{g}}$ on the Bloch sphere, 
forming the corner points $(-1)^g\Big((-1)^u,(-1)^{u+v},(-1)^v\Big)/\sqrt{3}$ of a cube. 
}
\label{fig:cube}
\end{center}
\end{figure}

Eight-state encoding was introduced in \cite{SdV2017} as a technique for 
quantum-encrypting classical data, to be used in e.g.\;Quantum Key Recycling (QKR) protocols.
The eight states are the QOTP encryptions of two specially chosen basis states:
\be
	\ket{\qj_0}\isdef \cos\fr\qa2 \ket0+ \sqrt i \sin\fr\qa2\ket1
	\quad , \quad
	\ket{\qj_1}\isdef \sin\fr\qa2 \ket0- \sqrt i \cos\fr\qa2\ket1
\ee
where $\qa\isdef\arccos\fr1{\sqrt3}$ and $\inprod{\qj_a}{\qj_b}=\qd_{ab}$.
On the Bloch sphere the $\ket{\qj_0}$ points in the direction $(1,1,1)$,
and $\ket{\qj_1}$ in the direction $(-1,-1,-1)$.
The QOTP-encrypted states are denoted as $\ket{u,v,g}$,
\be
	\ket{u,v,g}\isdef E_{uv}\ket{\qj_g} = \qs_z^u \qs_x^v \ket{\qj_g}.
\ee
Their position on the Bloch sphere is depicted in Fig.\;\ref{fig:cube}.

We label the four bases as $(u,v)$.
Instead of introducing four shadings or colors, we choose for a diagrammatic representation with labels,
\be
	\ket{u,v,i}\equiv \begin{tikzpicture}
		\begin{pgfonlayer}{nodelayer}
			\node [style=none] (2) at (0, 1) {};
			\node [style=none] (3) at (0, 1) {};
			\node [style=WideStateRight] (4) at (0, 0) {\rotatebox{90}{${}\!\!\!u,\!v,\!i$}};
			\node [style=none] (5) at (0, -0.25) {};
		\end{pgfonlayer}
		\begin{pgfonlayer}{edgelayer}
			\draw (3.center) to (5.center);
		\end{pgfonlayer}
	\end{tikzpicture}
	\quad\quad\quad\quad
	\sum_i \ket{u,v,i}^{\otimes n}\bra{u,v,i}^{\otimes m}\equiv \begin{tikzpicture}
		\begin{pgfonlayer}{nodelayer}
			\node [style=WhiteSpider] (0) at (0, 0) {\small $u,\!v$};
			\node [style=none] (1) at (-1, 1) {};
			\node [style=none] (2) at (-0.5, 1) {};
			\node [style=none] (3) at (1, 1) {};
			\node [style=none] (4) at (-1, -1.25) {};
			\node [style=none] (5) at (-0.5, -1.25) {};
			\node [style=none] (6) at (1, -1.25) {};
			\node [style=none] (7) at (-1, 1) {};
			\node [style=none] (8) at (0, -1.25) {};
			\node [style=none] (9) at (0.5, -1.25) {};
			\node [style=none] (10) at (0, 1) {};
			\node [style=none] (11) at (0.5, 1) {};
			\node [style=none] (12) at (0, -1.75) {\small $m$ inputs};
			\node [style=none] (13) at (0, 1.5) {\small $n$ outputs};
		\end{pgfonlayer}
		\begin{pgfonlayer}{edgelayer}
			\draw [bend left] (0) to (7.center);
			\draw [bend left=15] (0) to (2.center);
			\draw [bend right] (0) to (3.center);
			\draw [bend right] (0) to (4.center);
			\draw [bend right=15] (0) to (5.center);
			\draw [bend left] (0) to (6.center);
			\draw [style=new edge style 0] (10.center) to (11.center);
			\draw [style=new edge style 0] (8.center) to (9.center);
		\end{pgfonlayer}
	\end{tikzpicture}
\ee
The basis states are not depicted as left-right symmetric, since 
there is no basis in which all these eight states have a representation with real-valued 
vector components.

\subsection{Security of eight-state encoding}
\label{sec:8security}

The diagram for encoding and decoding of a classical bit, plus Eve's attack, is essentially the diagram for QOTP
of general states (\ref{equation:QOTPEve}), but now the input is a classical bit encoded in the $\qj_0,\qj_1$ basis.
\be
	\begin{tikzpicture}
		\begin{pgfonlayer}{nodelayer}
			\node [style=GraySpider] (0) at (-2.5, -4.25) {};
			\node [style=WHITESPIDER] (1) at (-3.25, -3.5) {};
			\node [style=none] (2) at (-3.25, -5.5) {};
			\node [style=none] (3) at (-3.25, -2.5) {};
			\node [style=GraySpider] (5) at (-1.75, -5.25) {};
			\node [style=none] (6) at (-3.25, -3.5) {};
			\node [style=GRAYSPIDER] (8) at (-3.25, -2.5) {};
			\node [style=WhiteSpider] (9) at (-2.5, -3.25) {};
			\node [style=WhiteSpider] (10) at (-1.75, -4.25) {};
			\node [style=GraySpider] (11) at (5, 3) {};
			\node [style=WHITESPIDER] (12) at (2.25, 4.25) {};
			\node [style=none] (13) at (2.25, 5) {};
			\node [style=none] (14) at (2.25, 2.5) {};
			\node [style=none] (18) at (-3.25, -2.5) {};
			\node [style=GRAYSPIDER] (19) at (2.25, 3.25) {};
			\node [style=WhiteSpider] (20) at (3.75, 2.75) {};
			\node [style=WhiteSpider] (21) at (-1.75, -4.25) {};
			\node [style=ASYM BOX] (23) at (0, 0.35) {$\Phi$};
			\node [style=none] (26) at (-3.25, -5) {};
			\node [style=none] (34) at (-3.25, -2) {};
			\node [style=none] (35) at (1.25, 1.5) {};
			\node [style=none] (36) at (0.5, 0.75) {};
			\node [style=none] (37) at (-2, -1) {};
			\node [style=none] (38) at (-0.25, 0.25) {};
			\node [style=none] (39) at (-0.25, 2) {};
			\node [style=none] (40) at (-0.25, 2.75) {};
			\node [style=none] (41) at (-1.325, -6.25) {};
			\node [style=none] (42) at (1.825, -6.25) {};
			\node [style=none] (43) at (1.825, 5) {};
			\node [style=none] (44) at (-1.325, 5) {};
			\node [style=none] (45) at (-2.25, -5.95) {\small Alice};
			\node [style=none] (46) at (0.1, -5.95) {\small Eve};
			\node [style=none] (47) at (2.625, -5.95) {\small Bob};
			\node [style=WIDESTATERIGHT] (50) at (-4, -5.5) {\;\footnotesize \rotatebox{90}{$\!\!00g\;\;\;$}};
			\node [style=WIDEEFFECTRIGHT] (60) at (2.25, 5.2) {\;\footnotesize \reflectbox{\rotatebox{90}{$\!\!00g'\;\;\;$}}};
		\end{pgfonlayer}
		\begin{pgfonlayer}{edgelayer}
			\draw [style=WIRE, bend left=45] (1) to (0);
			\draw [style=WIRE] (3.center) to (2.center);
			\draw (0) to (5);
			\draw [style=WIRE, bend left=45] (8) to (9);
			\draw (9) to (10);
			\draw [style=WIRE, bend left] (12) to (11);
			\draw [style=WIRE] (14.center) to (13.center);
			\draw [style=WIRE, bend left=45] (19) to (20);
			\draw [bend left=15] (20) to (21);
			\draw [bend left=15] (11) to (5);
			\draw [style=WIRE] (34.center) to (18.center);
			\draw [style=WIRE, bend left=45] (34.center) to (37.center);
			\draw [style=WIRE, bend right] (37.center) to (38.center);
			\draw [style=WIRE, bend left, looseness=1.25] (36.center) to (35.center);
			\draw [style=WIRE, bend right=45] (35.center) to (14.center);
			\draw [style=WIRE] (39.center) to (38.center);
			\draw [style=new edge style 0] (44.center) to (41.center);
			\draw [style=new edge style 0] (43.center) to (42.center);
		\end{pgfonlayer}
	\end{tikzpicture}
	\quad\quad = \quad\quad \sum_{uv} \begin{tikzpicture}
		\begin{pgfonlayer}{nodelayer}
			\node [style=none] (2) at (-3.25, -5.5) {};
			\node [style=none] (3) at (-3.25, -2.5) {};
			\node [style=none] (6) at (-3.25, -3.5) {};
			\node [style=none] (13) at (2.25, 5) {};
			\node [style=none] (14) at (2.25, 2.5) {};
			\node [style=none] (18) at (-3.25, -2.5) {};
			\node [style=ASYM BOX] (23) at (0, 0.35) {$\Phi$};
			\node [style=none] (26) at (-3.25, -5) {};
			\node [style=none] (34) at (-3.25, -2) {};
			\node [style=none] (35) at (1.25, 1.5) {};
			\node [style=none] (36) at (0.5, 0.75) {};
			\node [style=none] (37) at (-2, -1) {};
			\node [style=none] (38) at (-0.25, 0.25) {};
			\node [style=none] (39) at (-0.25, 2) {};
			\node [style=none] (40) at (-0.25, 2.75) {};
			\node [style=none] (41) at (-1.325, -3) {};
			\node [style=none] (42) at (1.825, -3) {};
			\node [style=none] (43) at ( 1.825, 3.4) {};
			\node [style=none] (44) at (-1.325, 3.4) {};
			\node [style=none] (45) at (-2.25, -3) {\small Alice};
			\node [style=none] (46) at (0.1, -3) {\small Eve};
			\node [style=none] (47) at (2.625, -3) {\small Bob};
			\node [style=WIDESTATERIGHT] (50) at (-4, -2.5) {\;\footnotesize \rotatebox{90}{$\!\!uvg\;\;\;$}};
			\node [style=WIDEEFFECTRIGHT] (60) at (2.25, 2.9) {\;\footnotesize \reflectbox{\rotatebox{90}{$\!\!uvg'\;\;\;$}}};
		\end{pgfonlayer}
		\begin{pgfonlayer}{edgelayer}
			\draw [style=WIRE] (34.center) to (18.center);
			\draw [style=WIRE, bend left=45] (34.center) to (37.center);
			\draw [style=WIRE, bend right] (37.center) to (38.center);
			\draw [style=WIRE, bend left, looseness=1.25] (36.center) to (35.center);
			\draw [style=WIRE, bend right=45] (35.center) to (14.center);
			\draw [style=WIRE] (39.center) to (38.center);
			\draw [style=new edge style 0] (44.center) to (41.center);
			\draw [style=new edge style 0] (43.center) to (42.center);
		\end{pgfonlayer}
	\end{tikzpicture}
\label{prot8state}
\ee
{\em After} the attack $\qF$, some information may leak about the basis $(u,v)$ or the data bit~$g$.
Here too it is important that a security property like (\ref{QKDsecurity}) holds \cite{LS2018}.
The main result of this paper is that indeed this is the case.

\begin{theorem}[Main result]
\label{th:main}
Consider the encoding of classical data as in diagram (\ref{prot8state}).
Let the label $uv$ in a spider denote the $(u,v)$-basis in 8-state encoding.
Non-disturbance in all four bases implies that Eve is decoupled from the Alice-to-Bob channel, 
\be
	\forall_{u,v\in\bits}\begin{tikzpicture}
		\begin{pgfonlayer}{nodelayer}
			\node [style=ASYM BOX] (0) at (0, 0) {$\Phi$};
			\node [style=none] (1) at (-0.75, 0) {};
			\node [style=none] (2) at (0, 0) {};
			\node [style=none] (3) at (-0.75, 1.25) {};
			\node [style=none] (4) at (-0.5, 1.25) {};
			\node [style=none] (5) at (0, -1.25) {};
			\node [style=WhiteSpider] (9) at (0.5, 1.25) {\tiny $uv$};
			\node [style=WhiteSpider] (10) at (0, -1.25) {\tiny $uv$};
			\node [style=none] (11) at (0.5, 2) {};
			\node [style=none] (12) at (0, -2) {};
			\node [style=none] (13) at (-0.25, 1.25) {};
			\node [style=none] (14) at (-1.25, 1.25) {};
			\node [style=none] (15) at (-0.5, 1.5) {};
			\node [style=none] (16) at (-1, 1.5) {};
			\node [style=none] (17) at (-0.65, 1.75) {};
			\node [style=none] (18) at (-0.65, 1.75) {};
			\node [style=none] (19) at (-0.85, 1.75) {};
			\node [style=none] (20) at (0.5, 0) {};
		\end{pgfonlayer}
		\begin{pgfonlayer}{edgelayer}
			\draw [style=WIRE] (3.center) to (1.center);
			\draw [style=WIRE] (2.center) to (5.center);
			\draw (11.center) to (9);
			\draw (10) to (12.center);
			\draw [style=WIRE] (18.center) to (19.center);
			\draw [style=WIRE] (15.center) to (16.center);
			\draw [style=WIRE] (13.center) to (14.center);
			\draw [style=WIRE] (9) to (20.center);
		\end{pgfonlayer}
	\end{tikzpicture} = \begin{tikzpicture}
	\begin{pgfonlayer}{nodelayer}
		\node [style=none] (0) at (0, 1) {};
		\node [style=none] (1) at (0, -1) {};
	\end{pgfonlayer}
	\begin{pgfonlayer}{edgelayer}
		\draw (0.center) to (1.center);
	\end{pgfonlayer}
\end{tikzpicture} 
	\quad\quad\quad\quad\implies\quad\quad\quad\quad
	\exists_\qr \begin{tikzpicture}
		\begin{pgfonlayer}{nodelayer}
			\node [style=ASYM BOX] (0) at (0, 0) {$\Phi$};
			\node [style=none] (1) at (0.5, 0) {};
			\node [style=none] (2) at (0, 0) {};
			\node [style=none] (3) at (0.5, 1.25) {};
			\node [style=none] (4) at (-0.5, 1.25) {};
			\node [style=none] (5) at (0, -1.25) {};
			\node [style=none] (6) at (-1, 1) {\tiny {\rm Eve}};
			\node [style=none] (7) at (1, 1) {\tiny {\rm Bob}};
			\node [style=none] (8) at (-0.75, -1) {\tiny {\rm Alice}};
			\node [style=none] (9) at (-0.5, 0) {};
		\end{pgfonlayer}
		\begin{pgfonlayer}{edgelayer}
			\draw [style=WIRE] (3.center) to (1.center);
			\draw [style=WIRE] (2.center) to (5.center);
			\draw [style=WIRE] (4.center) to (9.center);
		\end{pgfonlayer}
	\end{tikzpicture} = \begin{tikzpicture}
	\begin{pgfonlayer}{nodelayer}
		\node [style=STATERIGHT] (0) at (0, 0) {\rotatebox{90}{$\rho$}};
		\node [style=none] (1) at (0, 1) {};
		\node [style=none] (2) at (0, -0.5) {};
		\node [style=none] (3) at (-0.1, 1.2) {\tiny {\rm Eve}};
	\end{pgfonlayer}
	\begin{pgfonlayer}{edgelayer}
		\draw [style=WIRE] (1.center) to (2.center);
	\end{pgfonlayer}
\end{tikzpicture}  ~~~ \begin{tikzpicture}
\begin{pgfonlayer}{nodelayer}
	\node [style=none] (0) at (0, 1) {};
	\node [style=none] (1) at (0, -1.25) {};
	\node [style=none] (2) at (0.6, -1.1) {\tiny {\rm Alice}};
	\node [style=none] (3) at (0.5, 0.9) {\tiny {\rm Bob}};
\end{pgfonlayer}
\begin{pgfonlayer}{edgelayer}
	\draw [style=WIRE] (0.center) to (1.center);
\end{pgfonlayer}
\end{tikzpicture}
\label{8statesec}
\ee
\end{theorem}
\underline{\it Proof:}
We follow an approach similar to the proof of Theorem~3.1 in \cite{KTW2017}.
The strategy is to make use of Lemma~\ref{lemma:pullthrough} by attaching a disconnect
to the wire that goes from $V$ to Bob, and then pull the disconnect through~$V$.
However, we cannot literally re-use the steps of \cite{KTW2017} since our four bases are not mutually unbiased.
Hence we cannot create a disconnect using (\ref{equation:ComplementarityPositive});
instead we will construct a disconnect using (\ref{disconnect8}).

We start by first showing that a QOTP encryption can be `pulled through" the purification of~$\qF$.
This is proven as follows.
From Lemma~\ref{lemma:existsUpsi} and the condition in the left-hand side of (\ref{8statesec}) we get
\be
	\forall_{a,b\in\bits}\exists_{\qj_{ab},{\rm unitary}\,U_{ab}} \quad
	\begin{tikzpicture}
		\begin{pgfonlayer}{nodelayer}
			\node [style=asym box] (0) at (0, 0) {\reflectbox{V}};
			\node [style=none] (1) at (0, -1.5) {};
			\node [style=none] (11) at (-0.5, 0) {};
			\node [style=none] (12) at (-0.5, 1.5) {};
			\node [style=none] (13) at (0.35, 0) {};
			\node [style=none] (14) at (0.35, 1.5) {};
		\end{pgfonlayer}
		\begin{pgfonlayer}{edgelayer}
			\draw (13.center) to (14.center);
			\draw (12.center) to (11.center);
			\draw (0) to (1);
		\end{pgfonlayer}
	\end{tikzpicture} =
	\begin{tikzpicture}
		\begin{pgfonlayer}{nodelayer}
			\node [style=none] (0) at (0.8, 2.5) {};
			\node [style=none] (1) at (0.8, -3) {};
			\node [style=WhiteSpider] (2) at (0.8, -1) {\small $ab$};
			\node [style=medium box] (3) at (-1.2, 0.25) {$U_{ab}$};
			\node [style=StateMid] (4) at (-1.625, -1.5) {\rotatebox{-45}{$\psi_{ab}$}};
			\node [style=none] (5) at (-1.95, 0.25) {};
			\node [style=none] (6) at (-1.875, 0.75) {};
			\node [style=none] (7) at (-1.2, 0.75) {};
			\node [style=none] (8) at (-0.7, 0.75) {};
			\node [style=none] (9) at (-0.45, 0.25) {};
			\node [style=none] (10) at (-1.375, -2) {};
			\node [style=none] (11) at (-1.925, 0.75) {};
			\node [style=none] (12) at (-1.7, -2) {};
			\node [style=none] (13) at (-0.5, 2) {};
			\node [style=none] (14) at (-1.25, 2) {};
			\node [style=none] (15) at (-2, 2) {};
			\node [style=none] (16) at (-0.5, 0.75) {};
			\node [style=none] (17) at (-1.25, 0.75) {};
			\node [style=none] (18) at (-2, 0.75) {};
			\node [style=WhiteSpider] (19) at (-2, 2) {\small $ab$};
			\node [style=WhiteSpider] (20) at (-0.5, 2) {\small $ab$};
		\end{pgfonlayer}
		\begin{pgfonlayer}{edgelayer}
			\draw (0.center) to (1.center);
			\draw [bend left=45, looseness=1.25] (2) to (8.center);
			\draw [bend left=15] (7.center) to (10.center);
			\draw [bend right=15] (11.center) to (12.center);
			\draw (15.center) to (18.center);
			\draw (14.center) to (17.center);
			\draw (13.center) to (16.center);
		\end{pgfonlayer}
	\end{tikzpicture}
\label{existUpsiab}
\ee
Writing the $ab$-spiders as encryptions of $00$-spiders using phase spiders as in (\ref{encspiders}) yields
\be
	\begin{tikzpicture}
		\begin{pgfonlayer}{nodelayer}
			\node [style=asym box] (0) at (0, 0) {\reflectbox{V}};
			\node [style=none] (1) at (0, -1.5) {};
			\node [style=none] (11) at (-0.5, 0) {};
			\node [style=none] (12) at (-0.5, 1.5) {};
			\node [style=none] (13) at (0.35, 0) {};
			\node [style=none] (14) at (0.35, 1.5) {};
		\end{pgfonlayer}
		\begin{pgfonlayer}{edgelayer}
			\draw (13.center) to (14.center);
			\draw (12.center) to (11.center);
			\draw (0) to (1);
		\end{pgfonlayer}
	\end{tikzpicture} =
	\begin{tikzpicture}
		\begin{pgfonlayer}{nodelayer}
			\node [style=none] (0) at (2, 4) {};
			\node [style=none] (1) at (2, -3) {};
			\node [style=WhiteSpider] (2) at (2, -0.8) {\small $00$};
			\node [style=medium box] (3) at (-1.2, 0.25) {$U_{ab}$};
			\node [style=StateMid] (4) at (-1.625, -1.5) {\rotatebox{-45}{$\psi_{ab}$}};
			\node [style=none] (5) at (-1.95, 0.25) {};
			\node [style=none] (6) at (-1.875, 0.75) {};
			\node [style=none] (7) at (-1.2, 0.75) {};
			\node [style=none] (8) at (-0.7, 0.75) {};
			\node [style=none] (9) at (-0.45, 0.25) {};
			\node [style=none] (10) at (-1.375, -2) {};
			\node [style=none] (11) at (-1.925, 0.75) {};
			\node [style=none] (12) at (-1.7, -2) {};
			\node [style=none] (13) at (-0.5, 4) {};
			\node [style=none] (14) at (-1.25, 4) {};
			\node [style=none] (15) at (-2, 4) {};
			\node [style=none] (16) at (-0.5, 0.75) {};
			\node [style=none] (17) at (-1.25, 0.75) {};
			\node [style=none] (18) at (-2, 0.75) {};
			\node [style=WhiteSpider] (19) at (-2, 4) {\small $00$};
			\node [style=WhiteSpider] (20) at (-0.5, 4) {\small $00$};
			\node [style=GraySpider] (30) at (2, 0.5) {\tiny $b\!\pi$};
			\node [style=WhiteSpider] (31) at (2, 1.2) {\tiny $a\!\pi$};
			\node [style=GraySpider] (32) at (2, -1.8) {\tiny $b\!\pi$};
			\node [style=WhiteSpider] (33) at (2, -2.5) {\tiny $a\!\pi$};
			\node [style=GraySpider] (34) at (-2, 2.2) {\tiny $b\!\pi$};
			\node [style=WhiteSpider] (35) at (-2, 1.5) {\tiny $a\!\pi$};
			\node [style=GraySpider] (36) at (-0.5, 2.2) {\tiny $b\!\pi$};
			\node [style=WhiteSpider] (37) at (-0.5, 1.5) {\tiny $a\!\pi$};
			\node [style=GraySpider] (38) at (1.1, -0.9) {\tiny $b\!\pi$};
			\node [style=WhiteSpider] (39) at (0.4, -0.8) {\tiny $a\!\pi$};
		\end{pgfonlayer}
		\begin{pgfonlayer}{edgelayer}
			\draw (0.center) to (1.center);
			\draw [bend left=45, looseness=1.25] (2) to (8.center);
			\draw [bend left=15] (7.center) to (10.center);
			\draw [bend right=15] (11.center) to (12.center);
			\draw (15.center) to (18.center);
			\draw (14.center) to (17.center);
			\draw (13.center) to (16.center);
		\end{pgfonlayer}
	\end{tikzpicture}
	\stackrel{(*)}{=}  \begin{tikzpicture}
		\begin{pgfonlayer}{nodelayer}
			\node [style=none] (0) at (2, 4) {};
			\node [style=none] (1) at (2, -3) {};
			\node [style=WhiteSpider] (2) at (2, -0.8) {\small $00$};
			\node [style=medium box] (3) at (-1.2, 0.25) {$U_{00}$};
			\node [style=StateMid] (4) at (-1.625, -1.5) {\rotatebox{-45}{$\psi_{00}$}};
			\node [style=none] (5) at (-1.95, 0.25) {};
			\node [style=none] (6) at (-1.875, 0.75) {};
			\node [style=none] (7) at (-1.2, 0.75) {};
			\node [style=none] (8) at (-0.7, 0.75) {};
			\node [style=none] (9) at (-0.45, 0.25) {};
			\node [style=none] (10) at (-1.375, -2) {};
			\node [style=none] (11) at (-1.925, 0.75) {};
			\node [style=none] (12) at (-1.7, -2) {};
			\node [style=none] (13) at (-0.5, 4) {};
			\node [style=none] (14) at (-1.25, 4) {};
			\node [style=none] (15) at (-2, 4) {};
			\node [style=none] (16) at (-0.5, 0.75) {};
			\node [style=none] (17) at (-1.25, 0.75) {};
			\node [style=none] (18) at (-2, 0.75) {};
			\node [style=WhiteSpider] (19) at (-2, 4) {\small $00$};
			\node [style=WhiteSpider] (20) at (-0.5, 4) {\small $00$};
			\node [style=GraySpider] (30) at (2, 0.5) {\tiny $b\!\pi$};
			\node [style=WhiteSpider] (31) at (2, 1.2) {\tiny $a\!\pi$};
			\node [style=GraySpider] (32) at (2, -1.8) {\tiny $b\!\pi$};
			\node [style=WhiteSpider] (33) at (2, -2.5) {\tiny $a\!\pi$};
		\end{pgfonlayer}
		\begin{pgfonlayer}{edgelayer}
			\draw (0.center) to (1.center);
			\draw [bend left=45, looseness=1.25] (2) to (8.center);
			\draw [bend left=15] (7.center) to (10.center);
			\draw [bend right=15] (11.center) to (12.center);
			\draw (15.center) to (18.center);
			\draw (14.center) to (17.center);
			\draw (13.center) to (16.center);
		\end{pgfonlayer}
	\end{tikzpicture}
	\stackrel{(\ref{existUpsiab})}{=}
	\begin{tikzpicture}
		\begin{pgfonlayer}{nodelayer}
			\node [style=asym box] (0) at (0, 0) {\reflectbox{V}};
			\node [style=none] (1) at (0, -3) {};
			\node [style=none] (11) at (-0.5, 0) {};
			\node [style=none] (12) at (-0.5, 3) {};
			\node [style=none] (13) at (0.35, 0) {};
			\node [style=none] (14) at (0.35, 3) {};
			\node [style=GraySpider] (30) at (0, -1.3) {\tiny $b\!\pi$};
			\node [style=WhiteSpider] (31) at (0, -2) {\tiny $a\!\pi$};
			\node [style=GraySpider] (32) at (0.35, 1.5) {\tiny $b\!\pi$};
			\node [style=WhiteSpider] (33) at (0.35, 2.2) {\tiny $a\!\pi$};
		\end{pgfonlayer}
		\begin{pgfonlayer}{edgelayer}
			\draw (13.center) to (14.center);
			\draw (12.center) to (11.center);
			\draw (0) to (1);
		\end{pgfonlayer}
	\end{tikzpicture}
\label{V2pootjes}
\ee
In the step (*) we used that pulling the $ab$-encryptions into the $U_{ab}$ results in the $U_{00},\qj_{00}$ combination.
From (\ref{V2pootjes}) it follows that
\be
	\begin{tikzpicture}
		\begin{pgfonlayer}{nodelayer}
			\node [style=asym box] (0) at (0, 0) {\reflectbox{V}};
			\node [style=none] (1) at (0, -3) {};
			\node [style=none] (11) at (-0.5, 0) {};
			\node [style=none] (12) at (-0.5, 3) {};
			\node [style=none] (13) at (0.35, 0) {};
			\node [style=none] (14) at (0.35, 3) {};
			\node [style=GraySpider] (32) at (0.35, 1.5) {\tiny $b\!\pi$};
			\node [style=WhiteSpider] (33) at (0.35, 2.2) {\tiny $a\!\pi$};
		\end{pgfonlayer}
		\begin{pgfonlayer}{edgelayer}
			\draw (13.center) to (14.center);
			\draw (12.center) to (11.center);
			\draw (0) to (1);
		\end{pgfonlayer}
	\end{tikzpicture}
	= \begin{tikzpicture}
		\begin{pgfonlayer}{nodelayer}
			\node [style=asym box] (0) at (0, 0) {\reflectbox{V}};
			\node [style=none] (1) at (0, -3) {};
			\node [style=none] (11) at (-0.5, 0) {};
			\node [style=none] (12) at (-0.5, 3) {};
			\node [style=none] (13) at (0.35, 0) {};
			\node [style=none] (14) at (0.35, 3) {};
			\node [style=GraySpider] (30) at (0, -2) {\tiny $b\!\pi$};
			\node [style=WhiteSpider] (31) at (0, -1.3) {\tiny $a\!\pi$};
		\end{pgfonlayer}
		\begin{pgfonlayer}{edgelayer}
			\draw (13.center) to (14.center);
			\draw (12.center) to (11.center);
			\draw (0) to (1);
		\end{pgfonlayer}
	\end{tikzpicture}
\label{encthrough}
\ee
i.e.~the encryption can be `pulled through' $V$. 
Now we write, for any $u,v\in\bits$
\be
	\begin{tikzpicture}
		\begin{pgfonlayer}{nodelayer}
			\node [style=asym box] (0) at (0, 0) {\reflectbox{V}};
			\node [style=none] (1) at (0, -1.5) {};
			\node [style=none] (11) at (-0.5, 0) {};
			\node [style=none] (12) at (-0.5, 1.5) {};
			\node [style=none] (13) at (0.35, 0) {};
			\node [style=none] (14) at (0.35, 1.5) {};
		\end{pgfonlayer}
		\begin{pgfonlayer}{edgelayer}
			\draw (13.center) to (14.center);
			\draw (12.center) to (11.center);
			\draw (0) to (1);
		\end{pgfonlayer}
	\end{tikzpicture} 
	\stackrel{(\ref{disconnect8})}{\propto}  
	\sum_{a,b\in\bits}\begin{tikzpicture}
		\begin{pgfonlayer}{nodelayer}
			\node [style=asym box] (0) at (-0.25, -2.25) {\reflectbox{V}};
			\node [style=none] (1) at (-0.75, -2.25) {};
			\node [style=none] (2) at (0, -2.25) {};
			\node [style=none] (3) at (0, -1) {};
			\node [style=none] (4) at (0, -3.5) {};
			\node [style=none] (5) at (-1.5, 0) {};
			\node [style=WhiteSpider] (6) at (0, -1) {\tiny $uv$};
			\node [style=WhiteSpider] (7) at (-0.5, -0.25) {\tiny $uv$};
			\node [style=none] (8) at (1.5, 0.5) {};
			\node [style=WhiteSpider] (9) at (-0.5, 1) {\tiny $ab$};
			\node [style=WhiteSpider] (10) at (-0.5, 2) {\tiny $ab$};
			\node [style=none] (13) at (-0.5, -0.25) {};
			\node [style=none] (18) at (-0.5, 3) {};
		\end{pgfonlayer}
		\begin{pgfonlayer}{edgelayer}
			\draw (3.center) to (4.center);
			\draw [bend right=15] (5.center) to (1.center);
			\draw [bend left] (6) to (7);
			\draw [bend left] (8.center) to (6);
			\draw (10) to (9);
			\draw [style=WIRE] (7.center) to (9.center);
		\end{pgfonlayer}
	\end{tikzpicture}
	\stackrel{{\rm Lemma}~\ref{lemma:pullthrough}}{=}
	\sum_{a,b\in\bits}\begin{tikzpicture}
		\begin{pgfonlayer}{nodelayer}
			\node [style=asym box] (0) at (-0.75, 1.5) {\reflectbox{V}};
			\node [style=none] (1) at (-1.25, 1.5) {};
			\node [style=none] (2) at (-0.5, 1.5) {};
			\node [style=none] (3) at (0, -2) {};
			\node [style=none] (4) at (0, -3.25) {};
			\node [style=none] (5) at (-1.75, 2.75) {};
			\node [style=WhiteSpider] (6) at (0, -2) {\tiny $uv$};
			\node [style=WhiteSpider] (7) at (-0.5, -0.75) {\tiny $uv$};
			\node [style=none] (8) at (1.5, -0.5) {};
			\node [style=WhiteSpider] (9) at (-0.5, 0.25) {\tiny $ab$};
			\node [style=WhiteSpider] (10) at (-0.5, 2.8) {\tiny $ab$};
			\node [style=none] (13) at (-0.5, -0.5) {};
			\node [style=none] (18) at (-0.5, 0.5) {};
		\end{pgfonlayer}
		\begin{pgfonlayer}{edgelayer}
			\draw (3.center) to (4.center);
			\draw [bend right=15] (5.center) to (1.center);
			\draw [bend left=15] (6) to (7);
			\draw [bend left] (8.center) to (6);
			\draw (10) to (9);
			\draw [style=WIRE] (7.center) to (9.center);
		\end{pgfonlayer}
	\end{tikzpicture}
	\stackrel{(\ref{encthrough})}{=}
	\begin{tikzpicture}
		\begin{pgfonlayer}{nodelayer}
			\node [style=asym box] (0) at (-0.75, 2.25) {\reflectbox{V}};
			\node [style=none] (1) at (-1.25, 2.25) {};
			\node [style=none] (2) at (-0.5, 2.25) {};
			\node [style=none] (3) at (0, -2.5) {};
			\node [style=none] (4) at (0, -3.75) {};
			\node [style=none] (5) at (-1.75, 3.5) {};
			\node [style=WhiteSpider] (6) at (0, -2.5) {\tiny $uv$};
			\node [style=WhiteSpider] (7) at (-0.5, -1.5) {\tiny $uv$};
			\node [style=none] (8) at (1.5, -1) {};
			\node [style=WhiteSpider] (9) at (-0.5, 1) {\tiny $00$};
			\node [style=WhiteSpider] (10) at (-0.5, 3.5) {\tiny $00$};
			\node [style=GraySpider] (11) at (0.25, -1) {};
			\node [style=WHITESPIDER] (12) at (-0.5, -0.5) {};
			\node [style=none] (13) at (-0.5, -1.25) {};
			\node [style=none] (18) at (-0.5, 1.25) {};
			\node [style=GRAYSPIDER] (19) at (-0.5, 0.25) {};
			\node [style=WhiteSpider] (20) at (0.75, -0.75) {};
		\end{pgfonlayer}
		\begin{pgfonlayer}{edgelayer}
			\draw (3.center) to (4.center);
			\draw [bend right=15] (5.center) to (1.center);
			\draw [bend left=15] (6) to (7);
			\draw [bend left] (8.center) to (6);
			\draw (10) to (9);
			\draw [style=WIRE, bend left=45] (12) to (11);
			\draw [style=WIRE, bend left=45] (19) to (20);
			\draw [style=WIRE] (18.center) to (19);
			\draw [style=WIRE] (19.center) to (12.center);
			\draw [style=WIRE] (12.center) to (13.center);
		\end{pgfonlayer}
	\end{tikzpicture}
	\stackrel{(\ref{disconnect8})}{\propto}
	\begin{tikzpicture}
		\begin{pgfonlayer}{nodelayer}
			\node [style=asym box] (0) at (-0.75, 2.25) {\reflectbox{V}};
			\node [style=none] (1) at (-1.25, 2.25) {};
			\node [style=none] (2) at (-0.5, 2.25) {};
			\node [style=none] (3) at (0, -2.9) {};
			\node [style=none] (4) at (0, -4.15) {};
			\node [style=none] (5) at (-1.75, 3.5) {};
			\node [style=WhiteSpider] (6) at (0, -3.1) {\tiny $uv$};
			\node [style=WhiteSpider] (7) at (-0.5, -2.2) {\tiny $uv$};
			\node [style=none] (8) at (1.5, -1) {};
			\node [style=WhiteSpider] (9) at (-0.5, 1) {\tiny $00$};
			\node [style=WhiteSpider] (10) at (-0.5, 3.5) {\tiny $00$};
			\node [style=none] (13) at (-0.5, -1.25) {};
			\node [style=none] (18) at (-0.5, 1.25) {};
			\node [style=none] (32) at (-0.5, -1.95) {};
			\node [style=none] (33) at (-0.5, 0.225) {};
			\node [style=none] (37) at (-0.5, 1.05) {};
			\node [style=none] (45) at (-1.125, 0.225) {};
			\node [style=none] (46) at (0.125, 0.225) {};
			\node [style=none] (47) at (-0.875, -0.025) {};
			\node [style=none] (48) at (-0.175, -0.025) {};
			\node [style=none] (49) at (-0.625, -0.275) {};
			\node [style=none] (50) at (-0.375,  -0.275) {};
			\node [style=none] (51) at (-0.5, -1.275) {};
			\node [style=none] (52) at (0.125, -1.275) {};
			\node [style=none] (53) at (-1.125, -1.275) {};
			\node [style=none] (54) at (-0.125, -1.025) {};
			\node [style=none] (55) at (-0.875, -1.025) {};
			\node [style=none] (56) at (-0.375, -0.775) {};
			\node [style=none] (57) at (-0.625, -0.775) {};
			
		\end{pgfonlayer}
		\begin{pgfonlayer}{edgelayer}
			\draw (3.center) to (4.center);
			\draw [bend right=15] (5.center) to (1.center);
			\draw [bend left=15] (6) to (7);
			\draw [bend left] (8.center) to (6);
			\draw (10) to (9);
			\draw [style=WIRE] (37.center) to (33.center);
			\draw [style=WIRE] (45.center) to (46.center);
			\draw [style=WIRE] (47.center) to (48.center);
			\draw [style=WIRE] (49.center) to (50.center);
			\draw [style=WIRE] (52.center) to (53.center);
			\draw [style=WIRE] (54.center) to (55.center);
			\draw [style=WIRE] (56.center) to (57.center);
			\draw [style=WIRE] (51.center) to (32.center);
		\end{pgfonlayer}
	\end{tikzpicture}
\label{sec8finalpart}
\ee
In the second diagram the two $ab$-spiders can be contracted; then the remaining $ab$-spider can be written 
as the $ab$-encryption of a 00-spider, which is turned into a disconnect by the $\sum_{ab}$.
Then, the two $uv$-spiders fuse and form the identity operation.
This explains why the second diagram is equivalent to the first.
In the transition from the 3rd to the 4th diagram, both $ab$-spiders are rewritten as the $ab$-encryption of
a 00-spider; then the topmost encryption operator is pulled through $V$ and cancels the $ab$-encryption just below $V$. 

Finally, in the last expression in (\ref{sec8finalpart})
the trace on the $uv$-spider disappears and the two $uv$-spiders fuse to form the identity operator on the wire
from Alice to Bob.
Doubling the whole diagram yields the last expression in (\ref{8statesec}), with $\qr$ being the doubling of $V$ acting
on a random input.
\hfill$\square$

The proof deviates from \cite{KTW2017} in the way in which the disconnect is constructed in the second diagram of~(\ref{sec8finalpart}).
Whereas the disconnect in \cite{KTW2017} is based on complementary bases (\ref{equation:ComplementarityPositive}),
we make use of QOTP encryption~(\ref{disconnect8}).

\section{Discussion}

Most of what is written in this document is extracted from \cite{Coecke2017} and \cite{KTW2017},
and rehashed into an introduction that we hope is appealing to physicists.
Because of the peculiarities of 8-state encoding, with basis states that do not allow for a
simultaneous real-valued representation, we have chosen to explicitly show arrows on the wires in many of the diagrams.

The only really new content is Section~\ref{sec:8security}, the diagrammatic proof that non-disturbance
in 8-state encoding implies decoupling of Eve from Alice and Bob.
The security of 8-state encoding is of course known, but the derivation with diagrams is novel.

Our diagrammatic treatment of 8-state encoding may be of interest for Symmetric Informationally Complete measurements (`SIC-POVM') for tomography.

\vskip2mm

We have covered only the noiseless case. 
The situation with channel noise can be dealt with in exactly the same way as BB84/6-state QKD \cite{KTW2017}.
Here it is important to note that the diagrammatic approach does not yield improved bounds
(in fact they are worse than state-of-the-art).
Instead of considering it as a numerical tool, we see diagrams as an 
intuitive visualisation aid that is simple to master yet powerful enough to 
derive security proofs.



\subsection*{Acknowledgements}
We thank Aleks Kissinger and Daan Leermakers for useful discussions.


\bibliographystyle{plain}
\bibliography{diagr_arxiv}

\begin{thebibliography}{1}

\bibitem{AMTdW2000}
A.~Ambainis, M.~Mosca, A.~Tapp, and R.~de~Wolf.
\newblock {Private quantum channels}.
\newblock In {\em 41st Annual IEEE Symposium on Foundations of Computer
  Science}, pages 547--–553, 2000.

\bibitem{BB84}
C.~H. Bennett and G.~Brassard.
\newblock Quantum cryptography: Public key distribution and coin tossing.
\newblock In {\em Proceedings of IEEE International Conference on Computers,
  Systems and Signal Processing}, volume 175, 1984.

\bibitem{Coecke2017}
B.~Coecke and A.~Kissinger.
\newblock {\em {Picturing Quantum Processes}}.
\newblock Cambridge University Press, Cambridge, 2017.

\bibitem{Leung2002}
D.W.\;Leung.
\newblock {Quantum Vernam cipher}.
\newblock {\em Quantum Information and Computation}, 2(1):14--34, 2002.

\bibitem{KTW2017}
A.~Kissinger, S.~Tull, and B.~Westerbaan.
\newblock {Picture-perfect Quantum Key Distribution}, 2017.
\newblock \url{http://arxiv.org/abs/1704.08668}.

\bibitem{LS2018}
D.~Leermakers and B.~\v{S}kori\'{c}.
\newblock {Optimal attacks on qubit-based Quantum Key Recycling}.
\newblock {\em Quantum Information Processing}, 17(3):57, 2018.

\bibitem{BR2003}
P.O.\;Boykin and V.\;Roychowdhury.
\newblock Optimal encryption of quantum bits.
\newblock {\em Phys. Rev. A}, 67(4):042317, 2003.

\bibitem{SdV2017}
B.~\v{S}kori\'{c} and M.~de~Vries.
\newblock {Quantum Key Recycling with 8-state encoding (The Quantum One-Time
  Pad is more interesting than we thought)}.
\newblock {\em Int. J. of Quantum Information}, 15(3):1750016, 2017.

\bibitem{WolffsThesis}
Z.~Wolffs.
\newblock {An introduction to a novel diagrammatic notation for quantum
  mechanics and its applications to some quantum cryptographic protocols},
  2020.
\newblock Bachelor thesis, Maastricht University, The Netherlands.

\end{thebibliography}

\end{document}